\documentclass[acmsmall]{acmart}
\usepackage[most]{tcolorbox} 
\usepackage{multirow}
\usepackage{booktabs}
\usepackage{subcaption}
\usepackage{tabularx}
\usepackage{booktabs}
\usepackage{xcolor}
\usepackage{ragged2e}

\usepackage{soul} % 关键：支持自动换行的高亮
\definecolor{highlight}{RGB}{255,255,0} % 黄色
\sethlcolor{highlight} % 设置高亮底色

\AtBeginDocument{%
  }

\setcopyright{acmlicensed}
\copyrightyear{2026}
\acmYear{2026}
\acmDOI{XXXXXXX.XXXXXXX}

\acmJournal{TOSEM}
\acmVolume{1}
\acmNumber{1}
\acmArticle{1}
\acmMonth{5}
\begin{document}

\title{Keep Evaluation Fair: Detecting Data Leakage in Code Generation Benchmarks via Membership Inference Attacks}


\author{Dongdong Zhao}
\email{zdd@whut.edu.cn}
\affiliation{%
  \institution{School of Computer Science and Artificial Intelligence, Wuhan University of Technology}
  \city{Wuhan}
  \state{Hubei}
  \country{China}
}

\author{Jian Chen}
\email{321559@whut.edu.cn}
\affiliation{%
  \institution{School of Computer Science and Artificial Intelligence, Wuhan University of Technology}
  \city{Wuhan}
  \state{Hubei}
  \country{China}
}

\author{Guancheng Lin}
\email{guanchlin4-c@my.cityu.edu.hk}
\affiliation{%
  \institution{Department of Computer Science, City University of Hong Kong}
  \city{Hong Kong}
  \country{China}
}

\author{Jianwen Xiang}
\email{jwxiang@whut.edu.cn}
\affiliation{%
  \institution{Engineering Research Center of Transportation Information and Safety (ERCTIS), MoE of China, Wuhan University of Technology}
  \city{Wuhan}
  \state{Hubei}
  \country{China}
}

\author{Jacky Wai Keung}
\email{jacky.keung@cityu.edu.hk}
\affiliation{%
  \institution{Department of Computer Science, City University of Hong Kong}
  \city{Hong Kong}
  \country{China}
}

\author{Xiao Yu}
\authornote{Corresponding Author: Xiao Yu.}
\email{xiao.yu@zju.edu.cn}
\affiliation{%
  \institution{State Key Laboratory of Blockchain and Data Security, Zhejiang University}
  \city{Hangzhou}
  \state{Zhejiang}
  \country{China}
}

\renewcommand{\shortauthors}{Dongdong Zhao et al.}

\begin{abstract}

To systematically evaluate code generation capabilities of Large Language Models (LLMs), numerous benchmarks have been developed. However, a critical concern is the potential leakage of benchmark data into LLM training sets, which can inflate model performance and undermine the validity of evaluations. To the best of our knowledge, DetectLeak is currently the only method specifically designed to detect leakage in code generation benchmarks. It relies solely on perplexity scores, assuming that samples with lower perplexity are more likely leaked. However, perplexity primarily reflects a model’s general familiarity with common patterns and performs poorly on complex or rare samples. Moreover, DetectLeak overlooks other important features such as code similarity, functional correctness, and semantic embeddings of generated code, resulting in limited detection accuracy. To address the issues, we propose CGMIA (Code-Generation-specific Membership Inference Attack), a novel approach designed to detect data leakage in code generation benchmarks. 
CGMIA uses a shadow model fine-tuned on a subset of benchmark samples to mimic the target model’s behavior, creating labeled data where fine-tuned samples are members and held-out samples are non-members. For each sample, it collects the input prompt, shadow model’s generated code, and the reference solution, extracting both expert features (e.g., CodeBLEU, edit distance, test pass rate, perplexity) and semantic features from CodeBERT embeddings. These features are combined in an integrated learning module to capture surface-level memorization signals and deep behavioral patterns, enabling the classifier to accurately predict whether a sample was in the target model’s training set. Extensive experiments on eight widely used code generation benchmarks show that CGMIA significantly outperforms eight existing membership inference methods in most cases, while also effectively detecting known leaked APPS samples in StarCoder-7B's training data.

\end{abstract}




\received{20 February 2007}
\received[revised]{12 March 2009}
\received[accepted]{5 June 2009}

\maketitle

\section{Introduction}

Large Language Models (LLMs) have recently demonstrated remarkable capabilities in code generation tasks, driving substantial progress in automated software development ~\cite{sun2024sifting,mu2024clarifygpt,kou2024large,yu2024practitioners}. 
%   Practitioners’ expectations on automated test generation 
These advancements are largely attributed to large-scale pre-training on corpora that include not only natural language but also abundant source code ~\cite{wu2025data,yu2024makes}. 
To systematically assess the code generation abilities of LLMs, a growing number of benchmarks have been proposed.  Early benchmarks like APPS ~\cite{APPS}, HumanEval~\cite{humaneval}, and MBXP~\cite{MBXP} focus on algorithmic and basic programming tasks, while more recent benchmarks (e.g. RealisticCodeBench~\cite{yu2025realisticcodebench}, CoderEval \cite{CoderEval}, EvoCodeBench~\cite{EvoCodeBench}, ClassEval~\cite{ClassEval}) target more complex scenarios such as repository-level function generation and class-level code generation. 
% RealisticCodeBench: Towards More Realistic Evaluation of Large Language Models for Code Generation

However, since LLMs are typically pre-trained on massive internet-crawled datasets (e.g., GitHub), publicly available benchmark datasets risk unintentional inclusion in training data.
Indeed, recent studies~\cite{LessLeak-Bench, RiddellNC24, BalloccuSLD24, MattonSATAHMVGG24} show many widely used code generation benchmarks suffer from varying degrees of data leakage. For example, Zhou et al.~\cite{LessLeak-Bench} discovered that 108 samples from the APPS benchmark ~\cite{APPS} were included in the pre-training data of certain LLMs such as StarCoder-7B ~\cite{Starcoder}. As a result, StarCoder-7B exhibited substantially inflated performance on these leaked samples, achieving a pass@1 score that was 4.9 times higher than on unconfirmed samples. Such leakage raises serious concerns regarding the validity of benchmark-based evaluations, as it becomes unclear whether a model’s strong performance on the benchmark stems from genuine generalization or from memorization of benchmark samples that were present in its training data. In extreme cases, model developers could even intentionally incorporate benchmark data into training to artificially boost leaderboard scores ~\cite{zheng2024cheating}.  The challenge is further intensified by the proprietary nature of many commercial LLMs, which do not disclose their training data. Consequently, 
detecting benchmark data leakage by directly comparing LLM training corpora with code generation benchmark datasets, as adopted in prior work ~\cite{LessLeak-Bench}, is practically impossible.  Therefore, there is a pressing need for model-based leakage detection approaches that operate independently of LLM training data access, enabling fair and trustworthy evaluation of code generation models. 

\subsection{Motivation} 

To address this issue, Membership Inference Attacks (MIA) have been proposed as a way to detect whether specific samples were part of a model’s training set. Existing MIA methods can be broadly categorized into two types: classifier-based and metric-based  approaches~\cite{wu2024rethinking}. Classifier-based methods involve training a binary classifier to distinguish between a model’s behavioral patterns on member samples (i.e., samples used during training) versus non-member samples (i.e., unseen samples)\cite{yang2024unveiling,al2024traces}. 
A widely adopted approach in this category is the shadow model technique proposed by Shokri et al. ~\cite{shokri2017membership}. Here, an attacker trains a shadow model designed to mimic the target model’s behavior. Since the attacker can control the shadow model’s training process, they know exactly which data was used to train it. As a result, they can collect outputs from the shadow model on both its training (member) and non-training (non-member) samples, thus creating a labeled dataset that pairs sample features with their membership status. This dataset is then used to train the membership inference classifier.
Once trained, the membership inference classifier takes as input features extracted from the target model’s output on a new sample. Leveraging patterns learned from the shadow model’s member and non-member outputs, it analyzes these features to infer whether the new sample is likely a member of the target model’s training set.

In the code domain, classifier-based MIA methods like Gotcha~\cite{gotcha}, CodeMI ~\cite{codemi}, and TraWiC ~\cite{trawic} have been developed to protect the intellectual property rights of code data by determining if a code snippet was in an LLM’s training set, and thus potentially used without authorization. 
These methods typically prompt LLMs to perform code completion or masked token prediction and then extract features from the model’s responses to train a classifier. Gotcha~\cite{gotcha} encodes (input code, completed code, ground-truth) triplets using CodeBERT embeddings; CodeMI~\cite{codemi} extends this by incorporating ranking-based probability vectors of outputs as features; and TraWiC~\cite{trawic} computes match rates between the model’s predicted and original tokens via exact and fuzzy matching. 

Metric-based methods, by contrast, avoid training an explicit attack model. For example, Buzzer~\cite{buzzer}, also aimed at protecting code copyrights, introduces perturbations (e.g., case changes) to the input code and compares changes in the model’s [CLS] embedding, based on the assumption that member samples will be more sensitive to such perturbations. However, due to the inherent robustness of LLMs to minor textual variations, this method often fails to reliably distinguish member from non-member samples. DetectLeak~\cite{LessLeak-Bench} relies on perplexity scores, operating under the assumption that memorized samples will yield lower perplexity. However, perplexity largely reflects a model’s familiarity with common patterns, and thus performs poorly on complex or infrequent samples. Experimental results demonstrated that DetectLeak’s detection accuracy was only around 40–50\%.

Among the existing approaches ~\cite{gotcha, codemi, trawic, buzzer, LessLeak-Bench}, DetectLeak ~\cite{LessLeak-Bench} is the only method specifically designed for detecting leakage in full-program code generation benchmarks. In contrast, Gotcha~\cite{gotcha}, CodeMI~\cite{codemi}, TraWiC~\cite{trawic}, and Buzzer~\cite{buzzer} focus on copyright protection rather than how leakage undermines benchmark fairness. Therefore, these methods target partial code completion (e.g., function snippets or masked tokens), while full-program benchmarks require evaluating functional correctness—a dimension underaddressed by prior work. When benchmark samples leak into training data, models often generate code with abnormally high correctness (e.g., pass rates), presenting a potent yet underutilized signal for leakage detection.

\subsection{Our Work and Contributions}

Considering the issues, we propose CGMIA, a \underline{C}ode-\underline{G}eneration-specific \underline{M}embership \underline{I}nference \underline{A}ttack designed to detect potential data leakage in code generation benchmarks. CGMIA constructs a membership inference classifier to determine whether the sample (containing code generation prompt and reference solution) in a code generation benchmark belongs to the training set of the target model. Since the target LLM is typically accessible through black-box queries, we employ a shadow modeling strategy to simulate its behavior. Specifically, we fine-tune a shadow model on a subset of benchmark samples to create a labeled training set for the membership inference classifier: samples used in fine-tuning are designated as members, while the remaining unseen samples serve as non-members. For each sample, we gather the input prompt, reference solution, and output generated by the shadow model to form triplets of (prompt, generated code, reference solution) for the training set of the classifier. Unlike prior methods~\cite{gotcha, codemi, trawic, buzzer, LessLeak-Bench} that rely solely on either expert features or semantic embeddings, CGMIA integrates both: it extracts expert-designed features such as CodeBLEU~\cite{ren2020codebleu}, edit distance ~\cite{yujian2007normalized}, perplexity~\cite{jelinek1977perplexity}, and test pass rate, and combines them with semantic features derived from CodeBERT~\cite{CodeBERT} embeddings of the generated and reference code. These heterogeneous features, after feature dimension alignment, are fused via an integrated learning module to capture both surface-level memorization signals and deep behavioral patterns. The resulting membership inference classifier is then employed to predict the likelihood that a benchmark sample was included in the training set of the target LLM.

We select four recently released LLMs—Qwen2.5-Coder (3B), CodeGemma (2B), DeepSeek-Coder (1.3B), and Phi-2 (2.7B) as both target and shadow models for our study. Additionally, we choose eight benchmarks ranging from fundamental programming algorithm problems to class-level and repository-level code generation tasks. We compare our method against eight baseline approaches, including domain-specific methods (DetectLeak, Gotcha, CodeMI), machine learning classifiers (Logistic Regression, Random Forest) trained on expert features, and three deep learning architectures (CNN, LSTM, Transformer) that automatically learn semantic features from shadow model-generated code. 
% Experimental results demonstrate that our method achieves improvements of 6\% to 77\% in Precision, 9\% to 320\% in Recall, 29\% to 1160\% in MCC, and 10\% to 187\% in AUC compared to all baseline methods. 
Experimental results show that CGMIA consistently outperforms the baseline methods across eight code generation benchmarks and four target models. Compared with the strongest baseline for each target model, CGMIA achieves average absolute gains of 0.05–0.16 in Precision, 0.06–0.11 in Recall, 0.17–0.26 in MCC, and 0.09–0.13 in AUC. 
Furthermore, CGMIA can identify over 65\% of the 108 known leaked APPS samples in StarCoder’s training data, which were first reported by Zhou et al. \cite{LessLeak-Bench}, confirming its practical utility. The main contributions of this work are as follows:

(1) Unlike existing code membership inference studies that primarily focus on code completion or masked prediction for copyright protection, to the best of our knowledge, this work is the first to leverage MIA for data leakage detection in code generation benchmarks.

(2) We propose CGMIA, which integrates code-generation-specific expert features and semantic embeddings to jointly capture surface-level memorization signals and deeper behavioral patterns for accurate detection of code generation benchmark contamination.

(3) We conduct extensive experiments on eight code generation benchmarks and four open-source LLMs, demonstrating that CGMIA consistently outperforms eight baseline methods while further analyzing feature contributions, shadow-model mismatch, class imbalance, and real-world known leakage detection on StarCoder-7B/APPS.

\subsection{Organization}

The rest of this paper is organized as follows. In Section \ref{sec:realted_work}, we introduce the related work on membership inference attacks and data leakage detection for code generation benchmarks. Section \ref{sec:methodology} presents the methodology of CGMIA, including the task formulation, feature extraction, and membership inference classifier training. Section \ref{sec:Experimental_Setup} describes the experimental setup, including the target and shadow models, benchmarks, baselines, and evaluation metrics. Section \ref{sec:results} reports the experimental results and analyzes the effectiveness, robustness, and practical applicability of CGMIA. Section \ref{sec:Threats_of_validity} discusses the potential threats to validity. Finally, Section \ref{sec:Conclusion} concludes the paper.

\section{Related Work}
\label{sec:realted_work}

% todo加上，NLP和代码领域都有关注到数据泄露问题，所以有人尝试通过MIA解决。

In the code domain, classifier-based MIA methods are typically designed to determine whether a given code snippet was present in the training set of an LLM, with a primary focus on safeguarding code intellectual property. To achieve this, they usually prompt the model to perform code completion or masked token prediction, and then compare the model’s outputs with the original ground-truth code to assess potential leakage.  (1) Gotcha ~\cite{gotcha} fine-tunes a shadow model on a known subset of the target model’s training data. The shadow model is then tasked with code completion, and the resulting triplets (input code, completed code, ground truth) are collected. These triplets are encoded using CodeBERT and used as input features for a neural classifier that predicts membership. (2) CodeMI ~\cite{codemi} extends Gotcha by employing multiple shadow models to better capture the target model’s behavior. It also focuses on the code completion task, converting the model’s outputs into ranked probability vectors and using these ranking-based features to train a neural membership classifier. (3) TraWiC ~\cite{trawic} preprocesses each code snippet by masking selected identifiers (e.g., variable names, function names, and strings), converting them into (prefix, MASK, suffix) format. The model’s predictions for the MASK positions are then compared with the original code using exact matching (for variable names and function names) and fuzzy matching (for strings). The resulting match rates are aggregated into a feature vector, which is used to train a random forest classifier for membership inference. However, these methods focus on partial code completion or masked token prediction and are not fully suited for leakage detection in full-program code generation benchmarks.

Metric-based membership inference approaches, on the other hand, avoid training an attack model. Instead, they rely on statistical properties of the model’s outputs to infer membership. In the general machine/deep learning literature, for example, Min-k\% Pro Attack ~\cite{shi2023detecting} uses output confidence scores, while Loss Attack~\cite{yeom2018privacy} examines the model’s loss (e.g., cross-entropy loss) on each input to determine membership status. In the code domain, several analogous methods have emerged. (1) Buzzer ~\cite{buzzer} introduces perturbations (e.g., case transformations) to the input code and compares the changes in the model’s [CLS] embedding before and after perturbation. The assumption is that member samples exhibit larger changes, while non-members remain more stable. However, due to the robustness of LLMs to minor perturbations such as case changes, the attack effectiveness is limited.  (2) DetectLeak ~\cite{LessLeak-Bench} is the only method specifically designed for detecting leakage in code generation benchmarks. It infers membership by measuring the perplexity of generated code. Samples with lower perplexity are considered more likely to have been memorized,  indicating their presence in the target LLM's training data. However, perplexity mainly reflects a model’s general familiarity with common patterns and performs poorly on complex or rare samples. Experimental results also show that DetectLeak’s detection accuracy is only around 40–50\%, even below that of random guessing. Furthermore, a key challenge for metric-based methods lies in selecting appropriate thresholds; improper threshold can lead to high false positive or false negative rates, undermining the effectiveness of leakage detection.

Moreover, the above-mentioned works ~\cite{gotcha, codemi, trawic, buzzer} often overlook a key feature of code generation benchmarks: the availability of test cases to evaluate functional correctness. Leaked samples tend to yield abnormally high pass rates ~\cite{LessLeak-Bench}, providing a strong signal for leakage detection. To address these gaps, we propose a novel MIA method tailored for code generation benchmarks. Our approach combines expert-designed features—such as code correctness, similarity to ground truth, and perplexity—with semantic features to enable more robust and accurate detection of data leakage in code generation benchmarks.

\section{Methodology}
\label{sec:methodology}

\subsection{Task Formulation}
\label{sec:task_formulation}

\begin{figure*}[!t]
    \centering
    \includegraphics[width=\textwidth]{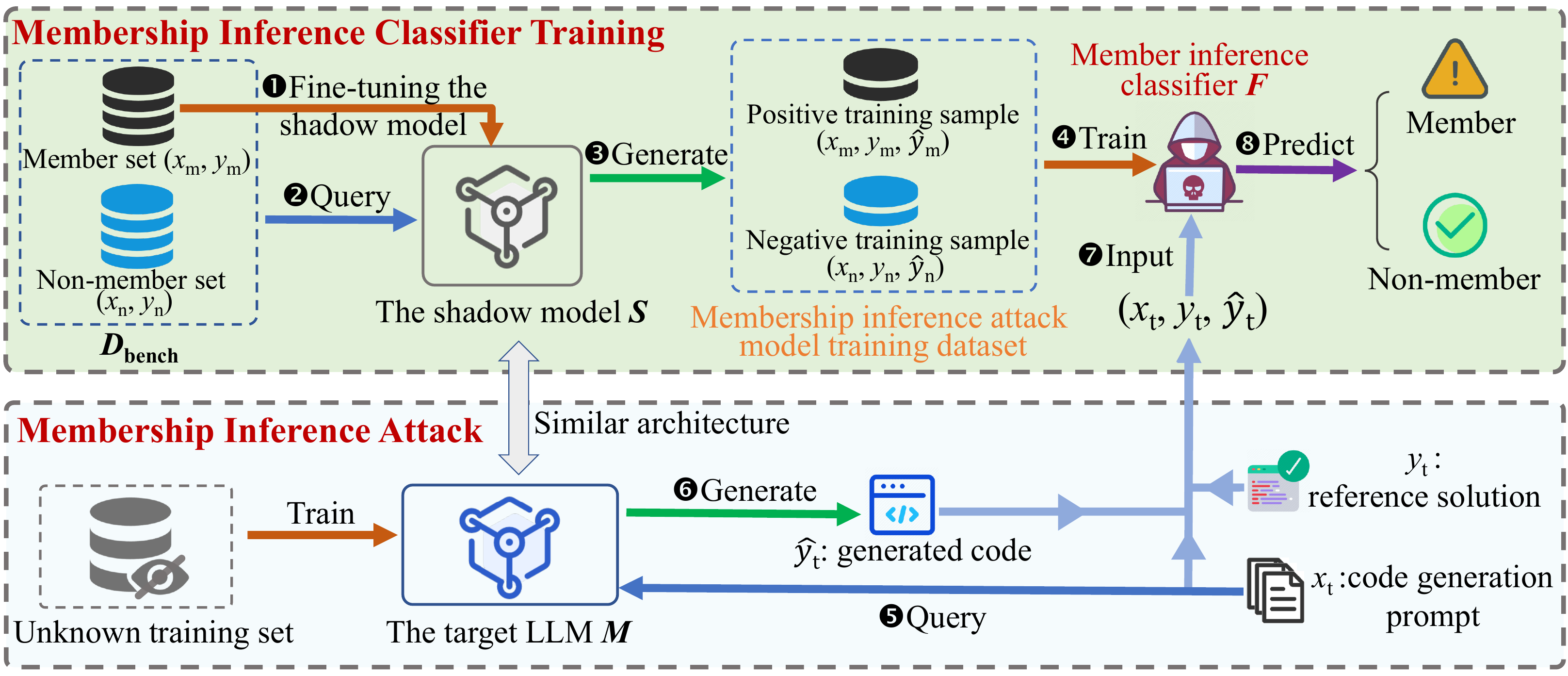}
    \caption{The overview of our CGMIA method.}
    \label{fig:framwork}
\end{figure*}

%Considering a target LLM $M$ trained or fine-tuned on a training set $D_{train}$, we assume that both the architecture and parameters of the model $M$ are unknown, and that the model can only be accessed via black-box queries (a common scenario in practice). That is, given an input $x$ (i.e., the code generation prompt), the model returns an output $\hat{y} = M(x)$ (i.e., the generated code). 

Considering a target LLM $M$ trained or fine-tuned on a training set $D_{train}$, we assume that the architecture, parameters, gradients, and training data of $M$ are inaccessible to the attacker, and that the model can only be queried through an external interface. 

More specifically, we adopt a query-based, score-access black-box setting: given an input $x$ (i.e., the code generation prompt), the attacker can obtain the generated output $\hat{y}=M(x)$ together with token-level log-probabilities for the generated sequence. These token-level log-probabilities are used to compute the perplexity feature in CGMIA. 
This score-access assumption is realistic in many practical deployment scenarios involving both commercial LLM services and open-weight LLMs whose training data are not publicly disclosed. For commercial or proprietary LLM services, several providers expose token-level log-probabilities through their APIs. For example, OpenAI supports \texttt{logprobs} and \texttt{top\_logprobs} in chat-completion interfaces \cite{openai_logprobs}, while Google Vertex AI provides \texttt{responseLogprobs} and \texttt{logprobs} for Gemini models \cite{google_vertexai_logprobs}. 
In addition, although many open-weight models publicly release their model weights, their training data are often partially disclosed or entirely unavailable. When these models are deployed through inference frameworks such as vLLM \cite{vllm_openai_compatible_server}, NVIDIA TensorRT-LLM \cite{tensorrt_llm_logprobs}, and llama.cpp \cite{llamacpp_logprobs}, the frameworks provide generation interfaces that expose token-level log-probabilities.

Under this practical score-access black-box scenario, a critical real-world concern arises regarding training data leakage in code evaluation benchmarks. 
In practice, the training data $D_{train}$ may unintentionally or even intentionally include samples from a code generation benchmark $D_{bench}$. Such data leakage can lead to artificially inflated performance on the benchmark, thereby undermining the reliability and fairness of evaluation results. To address this issue, we propose CGMIA,  a code-generation-specific MIA approach designed to detect potential data leakage in code generation benchmarks. The objective is to train a membership inference classifier $F$ designed to perform membership inference: given a prompt $x$, its reference solution $y$, and the generated output code $\hat{y} = M(x)$, the classifier predicts whether the sample $(x, y)$ is part of the model’s training set $D_{\text{train}}$: 

\begin{equation}
F(x, y, \hat{y})=\left
\{\begin{array}{rcl}
1  &   & {if~(x,y)\in D_{train}}  \\
0  & & {otherwise.}
\end{array}
\right. 
\end{equation}

Since $M$ is typically a black-box model, we adopt a shadow modeling approach widely used in MIA literature ~\cite{hu2022membership,hidano2021transmia,yan2022membership,shokri2017membership}, by training a shadow model with a known architecture and controllable training set to simulate $M$’s output behavior on member versus non-member samples.  As shown in Figure \ref{fig:framwork}, this process consists of two key stages. 

(1) \textbf{Membership Inference Classifier Training}: We begin by randomly sampling $p$ instances from the benchmark dataset $D_{\text{bench}}$ and using them to fine-tune the shadow model, denoted as $S$. These $p$ samples form the member set. For each $(x_m, y_m)$ in this set, we query $S$ to obtain its generated output $\hat{y}_m = S(x_m)$. Each resulting triplet $(x_m, y_m, \hat{y}_m)$ serves as a positive training example for the membership inference classifier $F$.
Next, we select the remaining $q$ samples from $D_{\text{bench}}$—those not used in the fine-tuning process. These comprise the non-member set. For each input $x_n$ in this set, we query $S$ to obtain $\hat{y}_n = S(x_n)$, producing negative training samples of the form $(x_n, y_n, \hat{y}_n)$ for $F$. To avoid issues related to data imbalance, we set 
$p$ equal to $q$. 
We then train the membership inference classifier $F$ on the combined set of positive and negative samples.

(2) \textbf{Membership Inference Attack}:
To detect potential benchmark leakage in the target model $M$, we query $M$ with benchmark samples $(x_t,y_t)$ $\in$ $D_{\text{bench}}$, 
and obtain outputs $\hat{y_t} = M(x_t)$. 
The resulting triplet $(x_t, y_t, \hat{y_t})$ 
is then fed into the trained classifier $F$, which predicts whether each sample is likely 
included in $M$’s training set $D_{train}$.

% \textcolor{blue}{Our experimental implementation follows this score-access setting. In our experiments, the target and shadow models are locally deployed LLMs, which allows us to obtain both generated code and token-level scores through the inference interface for computing perplexity. However, CGMIA does not use model parameters, gradients, internal activations, or training data at inference time. The local deployment is used only to emulate a practical score-access query interface and to make the experiments reproducible. Therefore, our setting should be interpreted as a score-access black-box setting rather than a strict text-only black-box setting.}

\subsection{Membership Inference Classifier Training}

The classifier training consists of two stages: feature extraction and integrated feature learning.  

\subsubsection{Feature Extraction}

Unlike existing MIA approaches such as Gotcha~\cite{gotcha}, which solely rely on semantic embeddings, or methods like DetectLeak~\cite{LessLeak-Bench}, CodeMI~\cite{codemi}, and TraWiC~\cite{trawic}, which focus exclusively on expert-designed features, our approach combines both expert features and semantic features. This hybrid strategy allows for a more comprehensive characterization of the model’s behavior, thereby enhancing the membership inference classifier’s ability to detect leakage. 

\textbf{Expert Features:}
Our selection of expert features is grounded in the hypothesis that if a sample in the code generation benchmark has been leaked into the training set and memorized by the target model, the generated output is more likely to exhibit memorization rather than generalization. Consequently, the generated code is expected to closely resemble the reference solution, and its probability of passing associated test cases will be higher \cite{salerno2025much}. 
This hypothesis is supported by multiple empirical studies. Yang et al. \cite{gotcha} demonstrated statistically significant differences in perplexity \cite{jelinek1977perplexity}, edit distance \cite{yujian2007normalized}, and BLEU scores \cite{papineni2002bleu} between samples successfully detected as leaked via Gotcha and non-leaked samples—validating that similarity-focused metrics can distinguish memorized outputs. Further support comes from Zhou et al. \cite{LessLeak-Bench}, who found that StarCoder-7B achieved a Pass@1 score on 108 detecting known leaked samples from the APPS benchmark 4.9 times higher than on unconfirmed samples, confirming that functional correctness metrics (e.g., Pass@k) also capture memorization signals.
These findings align with broader conventions in the code generation domain: Pass@k is widely used to evaluate the functional correctness of generated code \cite{humaneval, MBXP, ClassEval, EvoCodeBench}, while metrics such as edit distance \cite{yujian2007normalized} and CodeBLEU \cite{ren2020codebleu} are commonly employed to assess the syntactic and semantic similarity between generated code and reference solutions \cite{feng2024complexcodeeval,wang2022execution,zheng2024well,huang2022execution}. 
Building on this hypothesis, empirical evidence, and domain norms, we select the four expert features:

(a) \textbf{CodeBLEU}~\cite{ren2020codebleu} measures the semantic and syntactic similarity between the generated code $\hat{y}$ and reference solution $y$. A higher CodeBLEU means closer alignment with the reference solution.

(b) \textbf{Edit Distance}~\cite{yujian2007normalized} measures the normalized token-level edit distance between $\hat{y}$ and $y$, calculated as the minimum number of single-token edits (insertions, deletions, or substitutions) required to transform $\hat{y}$ to $y$. A lower edit distance implies that the generated output closely resembles the reference solution. 

% \textcolor{blue}{(c) \textbf{Perplexity}~\cite{jelinek1977perplexity} measures the likelihood of the generated output $\hat{y}$ given the input prompt $x$. Under the score-access black-box setting adopted in this work, perplexity is computed from the token-level log-probabilities assigned by the queried model to the generated sequence. A lower perplexity value suggests that the output is more predictable or familiar to the model, which may be indicative of memorized content. }

(c) \textbf{Perplexity}~\cite{jelinek1977perplexity} measures how likely the generated output
$\hat{y}$ is under the queried model, conditioned on the input prompt $x$. Under the
score-access black-box setting adopted in this work, the queried model returns token-level
log-probabilities for the generated sequence. Specifically, after tokenizing the generated
output $\hat{y}$ into a sequence of tokens $(\tau_1,\tau_2,\ldots,\tau_T)$, we compute
perplexity as:
\[
\operatorname{PPL}(\hat{y}\mid x)
=
\exp\left(
-\frac{1}{T}
\sum_{t=1}^{T}
\log p_M(\tau_t \mid x, \tau_1,\tau_2,\ldots,\tau_{t-1})
\right)
\]
where $T$ is the number of generated tokens, and
$p_M(\tau_t \mid x, \tau_1,\tau_2,\ldots,\tau_{t-1})$ denotes the probability assigned
by the queried model to the $t$-th generated token $\tau_t$, conditioned on the input
prompt $x$ and the previously generated tokens $\tau_1,\tau_2,\ldots,\tau_{t-1}$.
A lower perplexity value indicates that the generated output is more predictable or familiar
to the model, which may suggest memorization of benchmark-related content.

% （c）困惑度 [21] 用于衡量根据输入提示生成的输出 yˆ 的可能性。在本研究采用的评分访问黑箱设置下，困惑度是根据所查询模型对生成序列所赋予的词级对数概率计算得出的。困惑度值越低，表明输出对模型而言越具有可预测性或越熟悉，这可能意味着其中包含的是已记忆的内容。因此，此功能需要获取输出层面的分数，而非仅仅依赖生成的文本。

(d) \textbf{Test Pass Rate} is defined as a binary value set to 1 if the generated code $\hat{y}$ passes all test cases, and 0 otherwise. 

Given the stochastic nature of LLMs, we mitigate generation variance by sampling five outputs $\hat{y}$ for each input prompt $x$. Each of the five generated outputs is independently evaluated to compute the four expert features, resulting in a 20-dimensional expert feature vector per benchmark sample.

\textbf{Semantic Features:} We extract semantic features from five generated outputs $\hat{y}$ and the reference solution $y$ using the pre-trained CodeBERT model~\cite{CodeBERT}, which has demonstrated superior performance in capturing the contextual and structural semantics of code and has been extensively employed in various software engineering tasks~\cite{ni2022best,le2024software,akshar2024codebert,jiang2024stagedvulbert}.  
Due to the input-length limitation of CodeBERT, each input sequence can contain at most 512 tokens. To avoid directly truncating long code sequences, we process them using a sliding-window strategy. Specifically, we tokenize each generated output and reference solution using the CodeBERT tokenizer. For sequences longer than the input limit, we split the tokenized code into overlapping chunks. Each chunk contains at most 510 code tokens, and the required special tokens are then added to form a valid CodeBERT input whose total length does not exceed 512 tokens. We adopt a stride of 256 tokens between adjacent chunks to preserve contextual continuity. All chunks produced by the sliding window are encoded. Each chunk is independently fed into CodeBERT, and the hidden state of the first special token, i.e., the \texttt{[CLS]}, is used as the chunk-level representation. For a code sequence split into multiple chunks, we average all chunk-level representations to obtain a single 768-dimensional sequence-level embedding. For code sequences within the input limit, we directly use the \texttt{[CLS]} as the sequence-level embedding. This process is applied uniformly to the five generated outputs and the reference solution, resulting in six 768-dimensional vectors for each benchmark sample.

For each benchmark sample $i$, let the five generated outputs be 
$\hat{y}_i^1, \hat{y}_i^2, \ldots, \hat{y}_i^5$. After CodeBERT encoding, we obtain five generated-code embeddings $G_i = [g_i^1; g_i^2; \ldots; g_i^5] \in \mathbb{R}^{5 \times 768}$, where $g_i^k \in \mathbb{R}^{768}$ denotes the sequence-level embedding of the $k$-th generated output. We also obtain a reference-code embedding $r_i \in \mathbb{R}^{768}$ for the reference solution $y_i$. The expert feature vector is denoted as $e_i \in \mathbb{R}^{20}$, which is formed by concatenating the four expert features computed on each of the five generated outputs. In our implementation, CodeBERT is used as a frozen feature extractor, and its parameters are not updated during membership inference classifier training.

\subsubsection{Integrated Feature Learning.}

After extracting semantic features $G_i$ and expert features $e_i$, CGMIA integrates them into a unified representation for membership inference. Directly concatenating these semantic embeddings with the 20-dimensional expert feature vector would introduce a substantial dimensional imbalance between the two feature groups. Following Ni et al.~\cite{ni2022best}, before feature fusion, we apply dimension-alignment transformations to aggregate the generated-code embeddings and project the expert features into a comparable representation space. Specifically, the stacked generated-code representation $G_i$ is then passed through a one-dimensional CNN-based compression module. This module aggregates the five generated-code embeddings and outputs a single 768-dimensional generated-code semantic representation:
\begin{equation}
h_i^{gen}
=
\mathrm{Conv1D}(\mathbf{G}_i),
\quad
h_i^{gen} \in \mathbb{R}^{768}, G_i \in \mathbb{R}^{5 \times 768}.
\end{equation}
This step summarizes the semantic information of the five generated outputs while keeping the representation dimension consistent with the CodeBERT embedding size.
The reference solution is encoded separately by CodeBERT, and its semantic representation is denoted as:
\begin{equation}
h_i^{ref} \in \mathbb{R}^{768}.
\end{equation}
In parallel, the expert features computed from the five generated outputs form a 20-dimensional vector. Before fusion, this vector is standardized and then mapped into a 2048-dimensional representation through a linear projection:
\begin{equation}
h_i^{exp}
=
W_{e}\tilde{e}_i + b_e,
\quad
h_i^{exp} \in \mathbb{R}^{2048},
\end{equation}
where $\tilde{\mathbf{e}}_i \in \mathbb{R}^{20}$ denotes the standardized expert feature vector. This projection is used to reduce the imbalance caused by directly concatenating low-dimensional expert features with high-dimensional semantic embeddings.
Finally, we concatenate the reference-solution representation, the compressed generated-code representation, and the projected expert-feature representation to form the fused vector $\mathbf{z}_i$:
\begin{equation}
\mathbf{z}_i =
[h_i^{ref}; h_i^{gen}; h_i^{exp}],
\quad
z_i \in \mathbb{R}^{3584}.
\end{equation}

The fused representation is then fed into a three-layer feed-forward membership inference classifier. The classifier maps the 3584-dimensional input vector to a 1024-dimensional hidden representation, then to a 512-dimensional hidden representation, and finally to two output logits corresponding to the non-member and member classes. The output logits are converted into class probabilities using the softmax function.
Let $p_{i,1}$ denote the predicted probability that sample $i$ belongs to the member class, and let $p_{i,0}$ denote the predicted probability that sample $i$ belongs to the non-member class. The ground-truth membership label is denoted as $y_i \in \{0,1\}$, where 1 indicates member and 0 indicates non-member. The classifier is trained using the cross-entropy loss:
\begin{equation}
\mathcal{L}
=
-\frac{1}{N}
\sum_{i=1}^{N}
\left[
y_i \cdot \log(p_{i,1})
+
(1-y_i) \cdot \log(p_{i,0})
\right],
\end{equation}
where $N$ is the number of training samples.

During training, the parameters of the CNN-based semantic compression module and the expert-feature linear projection layer are frozen after initialization. Specifically, the CNN parameters used to aggregate the five generated-code embeddings, as well as the linear projection parameters $\mathbf{W}_{e}$ and $\mathbf{b}_{e}$ used for expert-feature dimension alignment, are not updated with membership labels. The remaining trainable modules are optimized using the shadow-model attack dataset and membership labels. This design is motivated by prior work ~\cite{rahimi2007random,dempster2020rocket} on fixed feature transformations and heterogeneous feature fusion. In machine learning, fixed feature transformations, such as random feature maps and random convolutional kernels~\cite{rahimi2007random,dempster2020rocket}, have been widely used to transform input data before training downstream classifiers, where the mappings or filters are not optimized using task labels but can still provide informative nonlinear representations while reducing the number of trainable parameters and the risk of overfitting. After training, the classifier predicts whether a benchmark sample is likely to be included in the target model's training set based on the fused semantic and expert features.

\section{Experimental Setup}
\label{sec:Experimental_Setup}

\subsection{Target Models and Shadow Models Implementations}

% 我们选择了CodeGemma(2B)\cite{codegemmateam2024codegemmaopencodemodels}, DeepSeek-Coder(1.3B), and Qwen2.5-Coder(3B) 这三个开源模型来作为目标模型和影子模型。主要的原因是他们有不错的代码生成能力。这些模型也常在软件工程论文中被使用。our study的目的是检验我们提出的CGMIA方法的有效性，并不是实际的要去检测某个llms的训练集是否包含了code generation benchmarks。特别是对于闭源llms，我们根本不知道他的训练集。 所以和这几篇论文的设置一样[1,2]，我们利用code generation benchmark中的一部分数据来微调一个llms，来模拟这个llms中存在code generation  benchmark sample泄露。举例的流程 我们按照如下的流程。
% The purpose of our study is to evaluate the effectiveness of the proposed CGMIA method, rather than to ascertain whether a specific LLM’s training set includes code generation benchmarks.

We select recent open-source models CodeGemma(2B)  \cite{codegemmateam2024codegemmaopencodemodels}, DeepSeek-Coder(1.3B) \cite{guo2024deepseek}, Qwen2.5-Coder(3B) \cite{qwen2.5}, and Phi-2(2.7B) \cite{phi-2} as our target and shadow models. 
Moreover, when these models were released, their pre-training and post-training datasets had already undergone deduplication, with benchmarks such as HumanEval and MBPP removed to ensure no leakage of benchmark data \cite{codegemmateam2024codegemmaopencodemodels,guo2024deepseek,qwen2.5,phi-2}.
The main reason is their strong code generation capabilities, and their frequent use in software engineering research ~\cite{xu2024cruxeval,lu2025next,yang2024swe,wang2024coderag}. 
% \textcolor{blue}{Similar to prior shadow-model-based membership inference attack studies in the code domain, such as Gotcha \cite{gotcha}, CodeMI \cite{codemi}, and AdvPrompt-MIA \cite{AdvPrompt-MIA}, we use a controlled fine-tuning procedure to construct an evaluation setting with known member and non-member labels. These studies commonly rely on controllable retraining or fine-tuning to build labeled attack datasets for quantitative evaluation. In this work, we follow the same general practice and fine-tune target and shadow models on selected benchmark subsets using LoRA \cite{lora}. Samples used for fine-tuning are treated as members, while held-out samples are treated as non-members.}
% Similar to the setups in previous studies~\cite{gotcha,codemi}, we utilize a portion of the data from the code generation benchmark to fine-tune an LLM using the LoRA method \cite{lora}, simulating the potential leakage of code generation benchmark samples within that LLM. 
For models that are either closed-source or open-source but do not fully disclose their training datasets, we can only rely on this simulated leakage approach.
Since the training data of these models remain hidden from researchers, even if CGMIA detects potential leakage by inferring the presence of specific benchmark samples in the training set, we are unable to verify these inferences due to the lack of access to the actual training data. Without ground-truth validation, we cannot confidently confirm the accuracy of CGMIA’s detections, which undermines our ability to evaluate its effectiveness in real-world scenarios. 
To implement this simulated leakage setup, the overall experimental procedure is outlined as follows: 

(1) To maintain black-box access to the target models, we ensure that the member and non-member datasets of the shadow models do not overlap with those of the target models. The code generation benchmark $D_{\text{bench}}$ is randomly and evenly partitioned into four mutually exclusive subsets: $D_{\text{bench1}}$, $D_{\text{bench2}}$, $D_{\text{bench3}}$, and $D_{\text{bench4}}$. 

(2) Target Model Implementations: For example, we select Qwen2.5-Coder as the target model and fine-tune it using  $D_{\text{bench1}}$.  The member data of the target model’s training set is $D_{\text{bench1}}$, while the non-member data is $D_{\text{bench2}}$.  

(3) Shadow Model Construction: For example, we select DeepSeek-Coder as the shadow model and fine-tune it using the subset $D_{\text{bench3}}$. We input $D_{\text{bench3}}$ into the fine-tuned shadow model and collect the input prompts \(x\), generated code \(\hat{y}\), and reference solutions \(y\); these serve as the positive training samples for the membership inference classifier \(F\). We then input $D_{\text{bench4}}$ into the fine-tuned shadow model, collect (\(x, \hat{y}, y\)) similarly; these serve as the negative training samples for the classifier \(F\). Finally, we train the membership inference classifier \(F\) using these positive and negative samples.

(4) Testing the Membership Inference Classifier: We input  $D_{\text{bench1}}$ and  $D_{\text{bench2}}$ into the target model Qwen-Coder to generate corresponding (\(x, \hat{y}, y\)). These triples are fed into the trained classifier \(F\), which predicts whether the data in  $D_{\text{bench1}}$ and  $D_{\text{bench2}}$ are member data of the target model.  

We use nucleus sampling with a top-p value of 0.95 and a temperature of 0.8, to prompt LLMs to produce diverse outputs. 
Due to the limited data volumes in some benchmarks, we train a single membership inference classifier using the combined training sets from all eight benchmarks, then individually predict which samples are leaked on each benchmark. To mitigate randomness, we repeat this process five times. The median of the five prediction results on this benchmark is considered the performance of the method on this benchmark.

\subsection{Implementation Details}

% \textcolor{blue}{To improve reproducibility, we provide the implementation details of the LoRA-based leakage simulation, feature construction, and membership inference classifier.}

% \textcolor{blue}{\textbf{LoRA-based leakage simulation.}
% We simulate benchmark leakage by fine-tuning the target and shadow models with LoRA. All models are loaded with half-precision floating-point format. The tokenizer uses the end-of-sequence token as the padding token, and right-side padding is adopted for causal language modeling. For LoRA fine-tuning, we set the rank $r$ to 8, the LoRA scaling factor $\alpha$ to 32, and the LoRA dropout rate to 0.04. The LoRA adapters are applied to the following target modules: \texttt{q\_proj}, \texttt{v\_proj}, \texttt{k\_proj}, \texttt{dense}, \texttt{fc1}, and \texttt{fc2}. We set \texttt{bias} to \texttt{none} and use \texttt{CAUSAL\_LM} as the task type.}

% We simulate benchmark leakage by fine-tuning the target and shadow models with LoRA. 
For all LoRA-based fine-tuning experiments, we use the same configuration: a rank of 8, a scaling factor $\alpha$ of 32, and a dropout rate of 0.04. The LoRA adapters are applied to the query, key, and value matrices in the attention modules, as well as the linear layers in the feed-forward networks. The adapters are trained using the AdamW optimizer with a learning rate of 2e-5 for 20 epochs. We set the fine-tuning batch size to 2 and the maximum sequence length to 2048 tokens. The detailed architectures of the compression module and the membership inference classifier are summarized in Table~\ref{tab:cnn_compression}, which specifies the parameter settings and input/output dimensions (where $B$ denotes the batch size). For training the membership inference classifier, we employ the Adam optimizer with a learning rate of 1e-4, a batch size of 100, and 50 epochs.

% \begin{table}[t]
% \centering
% \caption{\textcolor{blue}{Architecture of the generated-code semantic compression module.}}
% \label{tab:cnn_compression}
% \small
% \begingroup
% \color{blue}
% \begin{tabular}{lllll}
% \toprule
% \textbf{Layer} & \textbf{Input Shape} & \textbf{Output Shape} & \textbf{Kernel / Stride / Padding} & \textbf{Activation} \\
% \midrule
% Input & $B \times 5 \times 768$ & $B \times 5 \times 768$ & -- & -- \\
% Transpose & $B \times 5 \times 768$ & $B \times 768 \times 5$ & -- & -- \\
% Conv1D-1 & $B \times 768 \times 5$ & $B \times 768 \times 5$ & $k=3$, $s=1$, $p=1$ & ReLU \\
% Conv1D-2 & $B \times 768 \times 5$ & $B \times 768 \times 5$ & $k=3$, $s=1$, $p=1$ & -- \\
% Adaptive AvgPool1D & $B \times 768 \times 5$ & $B \times 768 \times 1$ & Output size = 1 & -- \\
% Squeeze & $B \times 768 \times 1$ & $B \times 768$ & -- & -- \\
% \bottomrule
% \end{tabular}
% \endgroup
% \end{table}

\begin{table}[t]
\centering
\caption{The architecture of the semantic compression module and membership inference classifier.}
\label{tab:cnn_compression}
\small
\begingroup
\begin{tabular}{lllll}
\toprule
\textbf{Layer} & \textbf{Input Shape} & \textbf{Output Shape} & \textbf{Parameter Setting} & \textbf{Activation} \\
\midrule
\multicolumn{5}{c}{\textbf{CNN architecture}} \\
\midrule
Input & $B \times 5 \times 768$ & $B \times 5 \times 768$ & -- & -- \\
Transpose & $B \times 5 \times 768$ & $B \times 768 \times 5$ & -- & -- \\
Conv1D-1 & $B \times 768 \times 5$ & $B \times 768 \times 5$ & $k=3$, $s=1$, $p=1$ & ReLU \\
Conv1D-2 & $B \times 768 \times 5$ & $B \times 768 \times 5$ & $k=3$, $s=1$, $p=1$ & -- \\
Adaptive AvgPool1D & $B \times 768 \times 5$ & $B \times 768 \times 1$ & Output size = 1 & -- \\
Squeeze & $B \times 768 \times 1$ & $B \times 768$ & -- & -- \\
\midrule
\multicolumn{5}{c}{\textbf{Membership inference classifier}} \\
\midrule
Input & $B \times 3584$ & $B \times 3584$ & -- & -- \\
FC-1 & $B \times 3584$ & $B \times 1024$ & -- & ReLU \\
FC-2 & $B \times 1024$ & $B \times 512$ & -- & ReLU \\
FC-3 & $B \times 512$ & $B \times 2$ & -- & -- \\
Softmax & $B \times 2$ & $B \times 2$ & -- & -- \\
\bottomrule
\end{tabular}
\endgroup
\end{table}

\subsection{Code Generation Benchmarks}
Considering the distribution of programming task difficulty, we select eight benchmark datasets for our experiments, as summarized in Table \ref{pass1}. HumanEval-X \cite{zheng2023codegeex} and MBXP \cite{MBXP} mainly consist of manually created algorithmic and basic programming tasks, focusing on fundamental coding abilities. NaturalCodeBench \cite{NaturalCodeBench} contains tasks derived from real and diverse code generation queries collected via the CodeGeeX~\cite{zheng2023codegeex} online service, reflecting more practical application scenarios. EvoCodeBench \cite{EvoCodeBench} targets repository-level code generation, with tasks collected from real GitHub Python repositories, representing more complex challenges. ClassEval \cite{ClassEval} features manually created class-level code generation tasks, designed to evaluate models’ understanding and generation of object-oriented Python programming structures. 
%Furthermore, Zhou et al. \cite{LessLeak-Bench} have not identified significant leakage of samples from these eight benchmarks into the training sets of mainstream LLMs, ensuring the reliability of our simulated leakage experiments. 
Among these benchmarks, HumanEval-X and MBXP offer multilingual versions. For consistency with other benchmarks and representativeness, we select Python and Java as the evaluation languages. The “Number” column in Table \ref{tab:datasets} indicates the total number of programming tasks in each benchmark.

The “Pass@1” column reports the pass@1 accuracy of four models—Qwen2.5-Coder, CodeGemma, DeepSeek-Coder, and Phi-2—on each dataset. For instance, on HumanEval-Python, their pass@1 scores are 48.7\%, 20.7\%, 59.7\%, and 30.5\%, respectively. Across all eight benchmarks, pass@1 values range from 1.4\% to 59.7\%, illustrating the wide variation in task difficulty. The “Fine-tuning Pass@1” column presents the performance of these models after fine-tuning on the benchmark datasets. It is evident that fine-tuning (i.e., with benchmark data leakage) substantially inflates the models’ performance, underscoring the impact of data leakage on evaluation results.

% \begin{table}[!t]
% \centering
% \caption{The details of the code generation benchmarks.}
% %\vspace{-5pt}
% \label{pass1}
% \resizebox{\linewidth}{!}{
% \footnotesize
% \begin{tabular}{c|c|c|c}
% \hline 
% Benchmark & Number & Pass@1 & Fine-tuning Pass@1 \\ 
% \hline
% HumanEval-X(Python) & 164 & 48.7/20.7/59.7/30.5 & 60.4/32.9/68.2/48.8\\ 
% \hline
% HumanEval-X(Java) & 164 & 59.6/23.6/45.3/29.3 & 65.8/36.6/55.3/47.6\\ 
% \hline
% MBXP-Python& 974 &56.2/33.1/46.2/35.9 & 62.8/40.6/51.4/49.3\\
% \hline
% MBXP-Java & 966 & 47.4/43.3/41.4/41.1 & 57.8/53.7/51.6/52.4\\
% \hline
% NaturalCodeBench-Python & 70 & 22.9/1.4/1.4/1.4 & 28.6/5.7/8.6/7.1\\
% \hline
% NaturalCodeBench-Java & 70 & 22.9/1.4/5.7/1.4 & 24.3/4.3/11.4/5.7\\
% \hline
% EvoCodeBench & 275 & 1.8/1.0/1.4/1.0 & 5.6/2.9/3.3/2.9/2.9\\
% \hline
% ClassEval & 100 & 24.0/14.0/15.0/15.0 & 34.0/21.0/27.0/22.0\\
% \hline
% \end{tabular}
% }
% \label{tab:datasets}
% \vspace{-20pt}
% \end{table}

\begin{table}[!t]
\centering
\caption{The details of the code generation benchmarks.}
\vspace{-0.3cm}
\label{pass1}
\footnotesize % 减小字体大小
\renewcommand{\arraystretch}{1.05} % 调整行高
\setlength{\tabcolsep}{4pt} % 减小列间距
\begin{tabular}{c|c|c|c}
\hline 
Benchmark & Number & Pass@1 & Fine-tuning Pass@1 \\ 
\hline
HumanEval-X(Python) & 164 & 48.7/20.7/59.7/30.5 & 60.4/32.9/68.2/48.8\\ 
\hline
HumanEval-X(Java) & 164 & 59.6/23.6/45.3/29.3 & 65.8/36.6/55.3/47.6\\ 
\hline
MBXP-Python& 974 &56.2/33.1/46.2/35.9 & 62.8/40.6/51.4/49.3\\
\hline
MBXP-Java & 966 & 47.4/43.3/41.4/41.1 & 57.8/53.7/51.6/52.4\\
\hline
NaturalCodeBench-Python & 70 & 22.9/1.4/1.4/1.4 & 28.6/5.7/8.6/7.1\\ % 缩写了长名称
\hline
NaturalCodeBench-Java & 70 & 22.9/1.4/5.7/1.4 & 24.3/4.3/11.4/5.7\\ % 缩写了长名称
\hline
EvoCodeBench & 275 & 1.8/0/1.4/1.0 & 5.6/2.9/3.3/2.9\\ % 修正了多余的数值
\hline
ClassEval & 100 & 24.0/14.0/15.0/15.0 & 34.0/21.0/27.0/22.0\\
\hline
\end{tabular}
\label{tab:datasets}
\vspace{-10pt}
\end{table}

\subsection{Baselines}
% This study draws inspiration from software defect detection methodologies. While some existing membership inference attack methods have been designed for code-related tasks, there remains a significant research gap in attack methods specifically targeting generative large language models. To address this, we build upon the foundational work of Chao Ni\cite{bl1} et al. and integrate approaches proposed by Zhou Yang et al.\cite{bl2} and Xin Zhou et al. \cite{bl3}to establish a benchmark framework for evaluating dataset leakage attacks in large language models. 

% 正如在section 2-相关工作中介绍的，在代码领域存在一些MIA methods。

As introduced in Section \ref{sec:realted_work} (Related Work), there are several MIA methods in the code domain. For baseline comparison, we select three recent methods: DetectLeak \cite{LessLeak-Bench} (proposed in 2025), and Gotcha \cite{gotcha} and CodeMI \cite{codemi} (both proposed in 2024) as our baselines. DetectLeak is specifically designed as an MIA method for code generation benchmarks, while both CodeMI and Gotcha are MIA approaches originally developed for enabling LLMs to perform code completion to investigate whether a code snippet exists in the LLMs’ training set. We adopt their methods as baselines, simply modifying the task from code completion to providing LLMs with code generation prompts, allowing the LLMs to generate code. 
We exclude TraWiC~\cite{trawic} and Buzzer~\cite{buzzer} from consideration because TraWiC’s masked token prediction approach and Buzzer’s code perturbation methodology render them incompatible with the code generation task. In our experimental setup, the test set consists of half membership samples; therefore, for the DetectLeak method, we designate the lower half of perplexity as membership samples. The parameter settings for Gotcha and CodeMI remain consistent with those in their original papers. 

In the field of machine/deep learning domains, although MIA research has been conducted since 2017 ~\cite{shokri2017membership}, most existing studies primarily focus on detecting whether samples are present in the training set of classification models, with a few addressing generative language models. These methods are not specifically designed for code models and are unsuitable for direct application to code generation tasks. Therefore, similar to Yang et al.~\cite{gotcha}, we adapt the existing approach proposed by Hisamoto et al.~\cite{hisamoto2020membership}, which was originally developed for natural language models, to serve as our baseline. This method employs a feature-based classification approach, manually extracting various features from model outputs and ground truth to construct a membership inference classifier for machine translation tasks. We modify their feature set to incorporate our four expert features (all feature values are normalized using min-max scaling). Following their experimental setup, which utilized multiple classifiers, our study employs several classifiers, including support vector machine, naive Bayes, decision tree, multi-layer perceptron, Logistic Regression (LR), and Random Forest (RF). Due to space limitations, in Section \ref{sec:results} (Experimental Results), we present only the results of the two best-performing machine learning classifiers: LR and RF. 
The  other machine learning baseline results are reported in Appendix.
Additionally, we consider three deep neural network architectures—CNN, LSTM, and Transformer—allowing them to automatically extract features from code generated by the shadow model and reference solution code without providing expert features.
Following existing work~\cite{tantithamthavorn2018impact, yu2023finding}, we also use grid search to tune the parameters of these machine learning classifiers. For the deep learning models (CNN, LSTM, and Transformer), the settings for epochs and learning rates are consistent with CGMIA. For all deep learning methods, we allocate 40\% of the training set as the validation set to prevent overfitting.

\subsection{Evaluation}
We employ $Precision = \frac{TP}{TP+FP}$, $Recall = \frac{TP}{TP+FN}$, $MCC=\frac{TP\times TN-FP\times FN}{\sqrt{(TP+FP)(TP+FN)(TN+FP)(TN+FN)}}$ as metrics, where TP denotes the number of member data correctly identified as members, FP represents the number of non-member data incorrectly classified as members, FN refers to the number of member data incorrectly predicted as non-members, and TN corresponds to the number of non-member data correctly predicted as such.  
In addition to Precision and Recall, we adopt the Matthews Correlation Coefficient (MCC) as a balanced metric for evaluating binary classification performance. Unlike Precision and Recall, which each reflect only specific aspects of the confusion matrix, MCC jointly considers TP, TN, FP, and FN, thereby providing a more comprehensive assessment of overall detection quality. The value of MCC ranges from $-1$ to $1$, where $1$ indicates perfect prediction, $0$ corresponds to performance equivalent to random guessing, and $-1$ indicates completely inverse prediction.
Additionally, we employ the Area Under the ROC Curve (AUC), which is computed as the area under the Receiver Operating Characteristic (ROC) curve. AUC measures the model's ability to distinguish between member and non-member samples across all possible classification thresholds, making it a threshold-independent evaluation metric. 
Thus, we utilize both MCC and AUC to assess the overall performance of our method from complementary perspectives under threshold-dependent and threshold-independent conditions.

We utilize the Wilcoxon signed-rank test \cite{Wilcoxon} and Cliff’s $\delta$ \cite{cliff} to assess the significance of the differences between our method and other baseline methods. To account for multiple comparisons, we apply the Benjamini-Hochberg (BH) procedure \cite{Haynes2013} to adjust the p-values. The null hypothesis of the Wilcoxon signed-rank test posits that there is no significant difference between the two methods, with a predefined significance level of 0.05 ($\alpha$ = 0.05). If the p-value after BH correction is less than 0.05, we reject the null hypothesis, indicating a statistically significant difference between the two approaches. In cases where the Wilcoxon signed-rank test reveals a significant difference, we employ Cliff's $\delta$ to determine the magnitude of the difference. The effect size is categorized as negligible (0 \textless  $\lvert$Cliff's $\delta$$\rvert$ \textless  0.147), small (0.147 $\leq$ $\lvert$Cliff's $\delta$$\rvert$ \textless  0.33), medium (0.33 $\leq$ $\lvert$Cliff's $\delta$$\rvert$ \textless 0.474), or large ($\lvert$Cliff's $\delta$$\rvert$  $\geq$ 0.474).
For statistical testing, each benchmark is treated as a paired observation. Specifically, we first compute Precision, Recall, MCC, and AUC over all test samples within each benchmark, and then compare CGMIA with each baseline method using the resulting benchmark-level metric values across the eight benchmarks through the Wilcoxon signed-rank test and Cliff's $\delta$.

\section{Experimental Results}
\label{sec:results}

We organize the results by addressing the Research Questions (RQs), which are presented as the titles for each section.

\subsection{\textbf{RQ1: Does CGMIA outperform other MIA methods?}}

Consistent with Yang et al.~\cite{gotcha}, in RQ1, we set the target model and the shadow model as the same LLMs. For example, we use the Qwen2.5-Coder fine-tuned on $D_{\text{bench1}}$ as the target model, and then use the Qwen2.5-Coder fine-tuned on $D_{\text{bench3}}$ as the shadow model. We input $D_{\text{bench3}}$ and $D_{\text{bench4}}$ into the shadow model and collect the triplets \((x, \hat{y}, y)\) as training data for the membership inference classifier \(F\). We then utilize \(F\) to predict whether the data in $D_{\text{bench1}}$ and $D_{\text{bench2}}$ are member data of the target model. In RQ3, we will use different shadow models to analyze the impact of the shadow model architecture on our CGMIA method. 
Table \ref{tab:precision}, Table \ref{tab:recall}, Table \ref{tab:mcc}, and Table \ref{tab:auc} present the Precision, Recall, MCC, and AUC of the MIA methods on the eight code generation benchmarks, respectively. In the tables, HP, HJ, MP, MJ, NP, NJ, EB, CE respectively represent the eight evaluation sets: HumanEval-X (Python), HumanEval-X (Java), MBXP-Python, MBXP-Java, NaturalCodeBench-Python, NaturalCodeBench-Java, EvoCodeBench, ClassEval; Go, DL, CI, Trans. respectively represent Gotcha, DetectLeak, CodeMI, Transformer; W/D/L indicates the number of datasets where our CGMIA method wins/draws/loses compared to baseline methods; Boldface indicates that the method achieves the best performance on this dataset.

As shown in Table \ref{tab:precision},  
our CGMIA method achieves the highest average Precision values of 0.89, 0.89, 0.92 and 0.82 across the eight benchmarks on the four models. Compared with the strongest baseline for each target model, CGMIA improves average Precision by 0.07, 0.05, 0.11, and 0.16 on Qwen2.5-Coder, CodeGemma, DeepSeek-Coder, and Phi-2, respectively.
The results show statistically significant superiority over most baseline methods, except for comparisons with DetectLeak and LSTM across Qwen2.5-Coder, CodeGemma, and DeepSeek-Coder, as well as occasional cases with Transformer and random forest. Among all baseline methods, random forest and CNN exhibit the second-best performance following our method. 

Table \ref{tab:recall} presents the Recall results across the four models. Our method achieves the highest Recall values of 0.89, 0.85, 0.83, and 0.84 across the eight benchmarks on the respective models. 
Compared with the strongest baseline for each target model, CGMIA improves average Recall by 0.11, 0.08, 0.06, and 0.11 on Qwen2.5-Coder, CodeGemma, DeepSeek-Coder, and Phi-2, respectively.
Statistical analysis using p-values and Cliff's $\delta$ demonstrates significant differences between our method and most baselines, with the exceptions of random forest and DetectLeak across Qwen2.5-Coder, CodeGemma, and DeepSeek-Coder, Gotcha on CodeGemma, and logistic regression on Phi-2. LSTM and CNN achieve the highest Recall performance after our method.

Tables \ref{tab:mcc} and \ref{tab:auc} present the MCC and AUC results across the eight benchmarks on the four models, which serve as comprehensive metrics for overall performance evaluation. Our method achieves the highest average MCC values of 0.75, 0.81, 0.73, and 0.63, and the highest average AUC scores of 0.95, 0.94, 0.95, and 0.89 across the eight benchmarks on the four models. 
Compared with the strongest baseline for each target model, CGMIA improves average MCC by 0.17, 0.26, 0.18, and 0.21, and improves average AUC by 0.10, 0.09, 0.09, and 0.13 on Qwen2.5-Coder, CodeGemma, DeepSeek-Coder, and Phi-2, respectively.
W/D/L shows our method outperforms the baselines on most benchmarks. Statistical analysis using p-values and Cliff's $\delta$ also indicates significant differences between our method and the baseline methods for both MCC and AUC metrics in most cases. Among the baseline methods, random forest and CNN demonstrate the best MCC and AUC performance after our method.

It is worth noting that, in this RQ, we follow the common MIA evaluation setting where member samples account for 50\% of the evaluation set. Accordingly, for DetectLeak, which infers membership based on the perplexity of generated code and treats samples with lower perplexity as more likely to have been memorized, we classify the 50\% samples with the lowest perplexity scores as members. However, in realistic deployment scenarios, the true member ratio is typically unknown in advance, making such threshold selection difficult in practice. 
Nevertheless, even under this favorable setting where the true member proportion is known, DetectLeak still performs substantially worse than CGMIA. Specifically, based on the average results across the eight benchmarks, CGMIA outperforms DetectLeak by 0.12--0.23 in Precision, 0.21--0.64 in Recall, 0.23--0.36 in MCC, and 0.18--0.58 in AUC across the four target models.  In contrast to DetectLeak, CGMIA adopts a classifier-based design that does not require manually specifying a prevalence-dependent membership threshold, while also achieving consistently better detection performance. Therefore, since DetectLeak still substantially underperforms CGMIA even under the favorable setting where the true member proportion is known, we focus our discussion on CGMIA as the primary practical solution for benchmark leakage detection under unknown-prevalence settings, and do not further investigate calibration strategies or abstention mechanisms specifically designed for DetectLeak.

In summary, across the four aforementioned tables, among the 128 total result sets (calculated as 8 baselines × 4 models × 4 metrics), 99 have p-values less than 0.05, while only 29 (23\%) have p-values greater than or equal to 0.05. This confirms that our method’s performance improvements over baseline methods are statistically significant in the vast majority of cases. In addition, the reference solutions in the eight code generation benchmarks contain between 152 and 1347 tokens. The results in RQ1 show that CGMIA’s performance does not exhibit any clear correlation with reference code length, indicating that solution length has little to no effect on detection performance.

\begin{table}[!t]
\centering
\caption{The Precision values of the methods.}
\vspace{-0.3cm}
\label{tab:precision}
\Huge  % 整体字体缩小
\renewcommand{\arraystretch}{1}  % 行间距加宽，提升可读性

% 第一行：Qwen2.5-Coder + CodeGemma
\begin{subtable}[t]{0.48\linewidth}
\centering
\caption{Qwen2.5-Coder}
\label{tab:precision_a}
\resizebox{1.05\linewidth}{!}{  % 自动适配宽度
\begin{tabular}{c|cccccccccccc}
        \hline
        \textbf{Bench} &\textbf{Go} &\textbf{DL} &\textbf{CI}  &\textbf{RF} &\textbf{LR} &\textbf{CNN} &\textbf{LSTM} &\textbf{Trans.} &\textbf{Ours} \\
        \hline
        
        HP & 0.63  & \textbf{1.00}  & 0.89  & 0.91  & 0.68  & 0.75  & 0.58  & 0.69  & 0.90  \\
        HJ & 0.80  & \textbf{0.97}  & 0.66  & 0.89  & 0.61  & 0.72  & 0.71  & 0.68  & 0.93  \\
        MP & 0.67  & 0.81  & 0.80  & 0.75  & 0.75  & 0.69  & 0.56  & 0.60  & \textbf{0.84}  \\
        MJ & 0.71  & 0.67  & 0.63  & 0.65  & 0.57  & 0.66  & 0.60  & 0.63  & \textbf{0.88}  \\
        NP & 0.80  & 0.00  & 0.56  & 0.80  & 0.49  & 0.90  & 0.83  & 0.79  & \textbf{1.00}  \\
        NJ & 0.96  & \textbf{1.00}  & 0.00  & 0.87  & 0.00  & 0.89  & 0.87  & 0.82  & 0.93  \\
        EB & 0.72  & \textbf{0.92}  & 0.73  & 0.89  & 0.53  & 0.77  & 0.86  & 0.69  & 0.79  \\
        CE & 0.89  & 0.75  & 0.61 & 0.74 & 0.58  & 0.82  & \textbf{0.99}  & 0.72  & 0.81  \\
        \hline
        AVG & 0.77  & 0.77  & 0.61  & 0.82  & 0.53  & 0.78  & 0.75  & 0.70  & \textbf{0.89} \\ \hline
        W/D/L & 6/0/2 & 4/0/4 & 8/0/0 & 6/0/2 & 8/0/0 & 7/0/1 & 6/0/2  & 8/0/0 & \\
        \hline
        p-value & \textbf{0.05} & 0.95 & \textbf{0.01} & 0.15 & \textbf{0.02} & \textbf{0.03} & 0.11 & 0.11 & \\
        \hline
        $\delta$ & \textbf{0.59} & \textbf{0.09} & \textbf{0.84} & \textbf{0.47} & \textbf{1.00} & \textbf{0.66} & \textbf{0.53} & \textbf{0.94} & \\
        \hline
        \end{tabular}
}
\end{subtable}
\hfill  % 两列间距自动填充
\begin{subtable}[t]{0.48\linewidth}
\centering
\caption{CodeGemma}
\label{tab:precision_b}
\resizebox{1.05\linewidth}{!}{
\begin{tabular}{c|cccccccccccc}
        \hline
        \textbf{Bench} &\textbf{Go} &\textbf{DL} &\textbf{CI}  &\textbf{RF} &\textbf{LR} &\textbf{CNN} &\textbf{LSTM} &\textbf{Trans.} &\textbf{Ours} \\
        \hline
        HP & 0.59  & \textbf{1.00}  & 0.72  & 0.81  & 0.67  & 0.79  & 0.66  & 0.70  & 0.94  \\
        HJ & 0.78  & \textbf{0.98}  & 0.77  & 0.94 & 0.64  & 0.74  & 0.70  & 0.70  & 0.96  \\
        MP & 0.65  & \textbf{0.90}  & 0.73  & 0.81  & 0.68  & 0.68  & 0.66  & 0.63  & 0.86  \\
        MJ & 0.71  & 0.81  & 0.65  & 0.77  & 0.60  & 0.66  & 0.60  & 0.64  & \textbf{0.89}  \\
        NP & 0.80  & 0.00  & 0.77  & 0.79  & 0.51  & \textbf{0.88}  & \textbf{0.88}  & 0.80  & 0.82  \\ 
        NJ & 0.96  & \textbf{1.00}  & 0.00  & 0.93  & 0.00  & 0.89  & 0.87  & 0.82  & 0.90  \\
        EB & 0.72  & 0.87  & 0.83  & 0.89  & 0.53  & 0.78  & \textbf{0.91}  & 0.71  & 0.85  \\
        CE & 0.87  & 0.60  & 0.75  & 0.85  & 0.59  & 0.82  & \textbf{0.99}  & 0.72  & 0.93  \\
        \hline
        AVG & 0.76  & 0.77  & 0.65  & 0.84  & 0.53  & 0.78  & 0.78  & 0.72  & \textbf{0.89} \\ \hline
        W/D/L & 7/0/0 & 3/0/5 & 8/0/0 & 6/0/2 & 8/0/0 & 7/0/1 & 5/0/3 & 8/0/0 & \\
        \hline
        p-value & \textbf{0.03} & 0.84 & \textbf{0.01} & 0.11 & \textbf{0.03} & \textbf{0.04} & 0.25 & \textbf{0.03} & \\
        \hline
        $\delta$ & \textbf{0.69} & 0.03 & \textbf{0.97} & \textbf{0.47} & \textbf{1.00} & \textbf{0.75} & \textbf{0.41} & \textbf{0.97} & \\
        \hline
        \end{tabular}
}
\end{subtable}

\vspace{1pt}  % 两行间距

% 第二行：DeepSeek-Coder + Phi-2（假设你补充对应真实数据）
\begin{subtable}[t]{0.48\linewidth}
\centering
\caption{DeepSeek-Coder}
\label{tab:precision_c}
\resizebox{1.05\linewidth}{!}{
\begin{tabular}{c|cccccccccccc}
        \hline
        \textbf{Bench} &\textbf{Go} &\textbf{DL} &\textbf{CI} &\textbf{RF} &\textbf{LR} &\textbf{CNN} &\textbf{LSTM} &\textbf{Trans.} &\textbf{Ours} \\
        \hline
        
        HP & 0.62  & \textbf{1.00}  & 0.83   & 0.93 & 0.67  & 0.78  & 0.66  & 0.67  & 0.92  \\
        HJ & 0.78  & \textbf{1.00}  & 0.64    & 0.96  & 0.57  & 0.74  & 0.69  & 0.69  & \textbf{1.00}  \\
        MP & 0.64  & 0.84  & 0.78    & 0.77  & 0.69  & 0.69  & 0.62  & 0.63  & \textbf{0.85}  \\
        MJ & 0.71  & 0.73  & 0.63    & 0.70  & 0.59  & 0.67  & 0.66  & 0.63  & \textbf{0.89}  \\
        NP & 0.81  & 0.00  & 0.72    & 0.71  & 0.51  & \textbf{0.86}  & 0.85  & 0.77  & 0.83  \\ 
        NJ & 0.95  & \textbf{1.00}  & 0.00    & 0.84  & 0.00  & 0.89  & 0.87  & 0.82  & \textbf{1.00}  \\
        EB & 0.72  & \textbf{0.88}  & 0.73   & 0.81 & 0.55  & 0.76  & 0.87  & 0.69  & 0.84  \\
        CE & 0.86  & 0.60  & 0.65    & 0.74  & 0.55  & 0.82  & 0.99  & 0.72  & \textbf{1.00}  \\
        \hline
        AVG & 0.76  & 0.76  & 0.62    & 0.81  & 0.52  & 0.78  & 0.77  & 0.70  & \textbf{0.92} \\ 
        \hline
        W/D/L & 8/0/0 & 4/2/2 & 8/0/0  & 7/0/1 & 8/0/0 & 7/0/1 & 6/0/2 & 8/0/0 & \\
        \hline
        p-value & \textbf{0.01} & 0.24 & \textbf{0.01}  & \textbf{0.02} & \textbf{0.03} & \textbf{0.04} & 0.05 & \textbf{0.03} & \\
        \hline
        $\delta$ & \textbf{0.75} & \textbf{0.23} & \textbf{1.00}& \textbf{0.63} & \textbf{1.00} & \textbf{0.81} & \textbf{0.59} & \textbf{1.00}  & \\
        \hline
        \end{tabular}
}
\end{subtable}
\hfill
\begin{subtable}[t]{0.48\linewidth}
\centering
\caption{Phi-2}
\label{tab:precision_d}
\resizebox{1.05\linewidth}{!}{
\begin{tabular}{c|cccccccccccc}
        \hline
        \textbf{Bench} &\textbf{Go} &\textbf{DL} &\textbf{CI} &\textbf{RF} &\textbf{LR} &\textbf{CNN} &\textbf{LSTM} &\textbf{Trans.} &\textbf{Ours} \\
        \hline
        
        HP & 0.57  & 0.53  & 0.50   & 0.55 & 0.76  & 0.62  & 0.48  & 0.53  & \textbf{0.96}  \\
        HJ & 0.58  & 0.59  & 0.67    & 0.50  & 0.69  & 0.62  & 0.56  & 0.58  & \textbf{0.90}  \\
        MP & 0.57  & 0.55  & 0.67    & 0.72  & 0.55  & 0.59  & 0.25  & 0.53  & \textbf{0.79}  \\
        MJ & 0.58  & 0.71  & 0.53    & 0.80  & 0.64  & 0.58  & 0.55  & 0.50  & \textbf{0.75}  \\
        NP & 0.63  & 0.58  & 0.45    & 0.50  & 0.69  & 0.60  & 0.64  & 0.56  & \textbf{0.65} \\ 
        NJ & 0.60  & 0.58  & 0.45    & 0.50  & \textbf{0.69}  & 0.60  & 0.64  & 0.56  & 0.65  \\
        EB & 0.58  & 0.52  & 0.58   & 0.79 & 0.62  & 0.62  & 0.62  & 0.53  & \textbf{0.85}  \\
        CE & 0.72  & 0.65  & 0.57    & 0.62  & 0.65  & 0.67  & \textbf{0.87}  & 0.48  & 0.80  \\
        \hline
        AVG & 0.61  & 0.59  & 0.56    & 0.64  & 0.66  & 0.62  & 0.59  & 0.50  & \textbf{0.82} \\ 
        \hline
        W/D/L & 8/0/0 & 8/0/0 & 8/0/0  & 7/0/1 & 7/0/1 & 8/0/0 & 7/0/1 & 8/0/0 & \\
        \hline
        p-value & \textbf{0.00} & \textbf{0.00} & \textbf{0.00}  & \textbf{0.02} & \textbf{0.00} & \textbf{0.00} & \textbf{0.02} & \textbf{0.00} & \\
        \hline
        $\delta$ & \textbf{0.21} & \textbf{0.23} & \textbf{0.26}& \textbf{0.18} & \textbf{0.16} & \textbf{0.20} & \textbf{0.23} & \textbf{0.32}  & \\
        \hline
        \end{tabular}
}
\end{subtable}
\vspace{-0.3cm}
\end{table}

\begin{table}[!t]
\centering
\caption{The Recall values of the methods.}
\vspace{-0.3cm}
\label{tab:recall}
\Huge  % 整体字体缩小
\renewcommand{\arraystretch}{1}  % 行间距加宽，提升可读性

% 第一行：Qwen2.5-Coder + CodeGemma
\begin{subtable}[t]{0.48\linewidth}
\centering
\caption{Qwen2.5-Coder}
\label{tab:recall_a}
\resizebox{1.05\linewidth}{!}{  % 自动适配宽度
\begin{tabular}{c|cccccccccccc}
        \hline
        \textbf{Bench} &\textbf{Go} &\textbf{DL} &\textbf{CI} &\textbf{RF} &\textbf{LR} &\textbf{CNN} &\textbf{LSTM} &\textbf{Trans.} &\textbf{Ours} \\
        \hline
        HP & 0.65  & 0.55  & 0.38  & 0.60  & 0.69  & 0.74  & 0.55  & 0.69  & \textbf{0.90}  \\
        HJ & 0.78  & \textbf{1.00}  & 0.73  & 0.99  & 0.83  & 0.72  & 0.59  & 0.67  & 0.84  \\
        MP & 0.68  & 0.74  & 0.46  & 0.77  & 0.61  & 0.69  & 0.51  & 0.60  & \textbf{0.88}  \\
        MJ & 0.70  & \textbf{0.98}  & 0.48  & 0.97  & 0.60  & 0.65  & 0.54  & 0.62  & 0.85  \\
        NP & 0.82  & 0.00  & 0.59  & 0.19  & 0.80  & \textbf{0.89}  & 0.81  & 0.78  & \textbf{0.89}  \\ 
        NJ & 0.96  & 0.38  & 0.00  & 0.71  & 0.00  & 0.88  & 0.85  & 0.82  & \textbf{1.00}  \\
        EB & 0.74  & \textbf{0.86}  & 0.50  & 0.79  & 0.75  & 0.76  & \textbf{0.86}  & 0.69  & 0.81  \\
        CE & 0.94  & 0.92  & 0.62  & 0.82 & 0.92  & 0.81  & \textbf{0.99}  & 0.72  & 0.93  \\
        \hline
        AVG & 0.78  & 0.68  & 0.47  & 0.73  & 0.65  & 0.77  & 0.71  & 0.70  & \textbf{0.89} \\ \hline
        W/D/L & 7/0/1 & 5/0/3 & 8/0/0 & 6/0/2 & 8/0/0 & 7/0/1 & 6/0/2 & 8/0/0 & \\
        \hline
        p-value & \textbf{0.02} & 0.31 & \textbf{0.01} & 0.25 & \textbf{0.02} & \textbf{0.04} & \textbf{0.04} & \textbf{0.02} &  \\
        \hline
        $\delta$ & \textbf{0.53} & \textbf{0.27} & \textbf{1.00} & \textbf{0.53} & \textbf{0.78} & \textbf{0.72} & \textbf{0.63} & \textbf{0.97} & \\
        \hline
        \end{tabular}
}
\end{subtable}
\hfill  % 两列间距自动填充
\begin{subtable}[t]{0.48\linewidth}
\centering
\caption{CodeGemma}
\label{tab:recall_b}
\resizebox{1.05\linewidth}{!}{
\begin{tabular}{c|cccccccccccc}
        \hline
        \textbf{Bench} &\textbf{Go} &\textbf{DL} &\textbf{CI}  &\textbf{RF} &\textbf{LR} &\textbf{CNN} &\textbf{LSTM} &\textbf{Trans.}&\textbf{Ours} \\
        \hline
        HP & 0.55  & 0.02  & 0.34  & 0.18  & 0.61  & 0.78  & 0.54  & 0.70  & \textbf{0.89}  \\
        HJ & 0.73  & 0.88  & 0.64  & \textbf{0.92}  & 0.75  & 0.71  & 0.59  & 0.69  & 0.84  \\
        MP & 0.66  & 0.27  & 0.38  & 0.48  & 0.55  & 0.67  & 0.50  & 0.63  & \textbf{0.87}  \\
        MJ & 0.72  & 0.84  & 0.51  & 0.84  & 0.67  & 0.66  & 0.53  & 0.64  & \textbf{0.85}  \\
        NP & 0.71  & 0.00  & 0.28  & 0.12  & 0.75  & \textbf{0.88}  & \textbf{0.88}  & 0.80  & 0.82  \\ 
        NJ & \textbf{0.97}  & 0.23  & 0.00  & 0.42  & 0.00  & 0.88  & 0.85  & 0.82  & 0.89  \\
        EB & 0.66  & \textbf{0.98}  & 0.23  & 0.92  & 0.64  & 0.76  & 0.91  & 0.71  & 0.85  \\
        CE & 0.92  & \textbf{1.00}  & 0.52  & 0.96 & 0.93  & 0.81  & 0.99  & 0.72 & 0.76  \\
        \hline
        AVG & 0.74  & 0.53  & 0.36  & 0.61  & 0.61  & 0.77  & 0.72  & 0.71  & \textbf{0.85} \\ \hline
        W/D/L & 6/0/2 & 5/0/3 & 8/0/0 & 5/0/3 & 7/0/1 & 6/0/2 & 5/0/3 & 8/0/0 & \\
        \hline
        p-value & 0.11 & 0.25 & \textbf{0.01} & 0.25 & \textbf{0.04} & \textbf{0.01} & 0.25 & \textbf{0.02} & \\
        \hline
        $\delta$ & \textbf{0.50} & \textbf{0.22} & \textbf{1.00} & \textbf{0.22} & \textbf{0.75} & \textbf{0.56} & \textbf{0.19} & \textbf{0.94} & \\
        \hline
        \end{tabular}
}
\end{subtable}

\vspace{1pt}  % 两行间距

% 第二行：DeepSeek-Coder + Phi-2（假设你补充对应真实数据）
\begin{subtable}[t]{0.48\linewidth}
\centering
\caption{DeepSeek-Coder}
\label{tab:recall_c}
\resizebox{1.05\linewidth}{!}{
\begin{tabular}{c|cccccccccccc}
        \hline
        \textbf{Bench} &\textbf{Go} &\textbf{DL} &\textbf{CI} &\textbf{RF} &\textbf{LR} &\textbf{CNN} &\textbf{LSTM} &\textbf{Trans.} &\textbf{Ours} \\
        \hline
        
        HP & 0.63  & 0.44  & 0.50  & 0.51  & 0.59  & \textbf{0.78}  & 0.55  & 0.67  & 0.76  \\
        HJ & 0.73  & \textbf{0.99}  & 0.59  & 0.99  & 0.71  & 0.72  & 0.58  & 0.67  & 0.82  \\
        MP & 0.66  & \textbf{0.92}  & 0.43  & 0.94 & 0.62  & 0.69  & 0.50  & 0.63  & 0.88  \\
        MJ & 0.73  & \textbf{0.98}  & 0.51  & 0.98  & 0.64  & 0.67  & 0.54  & 0.62  & 0.86  \\
        NP & 0.75  & 0.00  & 0.34  & 0.17  & 0.80  & \textbf{0.86}  & 0.84  & 0.77  & 0.77  \\ 
        NJ & \textbf{0.97}  & 0.05  & 0.00  & 0.18  & 0.00  & 0.88  & 0.85  & 0.82  & 0.93  \\
        EB & 0.72  & 0.60  & 0.21  & 0.55  & 0.62  & 0.75  & \textbf{0.87}  & 0.69  & 0.81  \\
        CE & 0.91  & 0.91  & 0.28  & 0.86  & 0.75  & 0.81  & \textbf{0.99}  & 0.72  & 0.78  \\
        \hline
        AVG & 0.76  & 0.61  & 0.36  & 0.65  & 0.59  & 0.77  & 0.72  & 0.70  & \textbf{0.83} \\ \hline
        W/D/L & 6/0/2 & 4/0/4 & 8/0/0 & 4/0/4 & 7/0/1 & 5/0/3 & 5/0/3 & 8/0/0 & \\
        \hline
        p-value & \textbf{0.02} & 0.31 & \textbf{0.01} & 0.31 & \textbf{0.01} & \textbf{0.02} & \textbf{0.04} & \textbf{0.01} & \\
        \hline
        $\delta$ & \textbf{0.53} & \textbf{0.27} & \textbf{1.00} & \textbf{0.09} & \textbf{0.78} & \textbf{0.72} & \textbf{0.63} & \textbf{0.97} & \\
        \hline
        \end{tabular}
}
\end{subtable}
\hfill
\begin{subtable}[t]{0.48\linewidth}
\centering
\caption{Phi-2}
\label{tab:recall_d}
\resizebox{1.05\linewidth}{!}{
\begin{tabular}{c|cccccccccccc}
        \hline
        \textbf{Bench} &\textbf{Go} &\textbf{DL} &\textbf{CI} &\textbf{RF} &\textbf{LR} &\textbf{CNN} &\textbf{LSTM} &\textbf{Trans.} &\textbf{Ours} \\
        \hline
        
        HP & 0.53  & 0.07  & 0.25  & 0.19  & \textbf{0.86}  & 0.60  & 0.50  & 0.53  & 0.83  \\
        HJ & 0.52  & 0.18  & 0.64  & 0.00  & \textbf{0.89}  & 0.61  & 0.54  & 0.57  & 0.82  \\
        MP & 0.58  & 0.11  & 0.47  & 0.66  & \textbf{0.86}  & 0.59  & 0.50  & 0.53  & 0.82 \\
        MJ & 0.60  & 0.49  & 0.50  & 0.83  & 0.67  & 0.58  & 0.52  & 0.50  & \textbf{0.93}  \\
        NP & 0.51  & 0.23  & 0.42  & 0.00  & 0.75  & 0.59  & 0.63  & 0.54  & \textbf{0.92}  \\ 
        NJ & 0.52  & 0.16  & 0.17  & 0.4  & 0.62  & 0.63  & 0.74  & 0.50  & \textbf{0.83}  \\
        EB & 0.58  & 0.04  & 0.47  & \textbf{0.82}  & 0.49  & 0.60  & 0.62  & 0.51  & \textbf{0.82}  \\
        CE & 0.76  & 0.30  & 0.76  & 0.72  & 0.68  & 0.67  & \textbf{0.87}  & 0.49  & 0.77  \\
        \hline
        AVG & 0.57  & 0.20  & 0.46  & 0.45  & 0.73  & 0.61  & 0.61  & 0.52  & \textbf{0.84} \\ \hline
        W/D/L & 8/0/0 & 8/0/0 & 8/0/0 & 7/1/0 & 5/0/3 & 8/0/0 & 7/0/1 & 8/0/0 & \\
        \hline
        p-value & \textbf{0.01} & \textbf{0.00} & \textbf{0.00} & \textbf{0.02} & 0.07 & \textbf{0.00} & \textbf{0.01} & \textbf{0.00} & \\
        \hline
        $\delta$ & \textbf{0.27} & \textbf{0.64} & \textbf{0.38} & \textbf{0.39} & \textbf{0.11} & \textbf{0.23} & \textbf{0.23} & \textbf{0.32} & \\
        \hline
        \end{tabular}
}
\end{subtable}
\vspace{-0.3cm}
\end{table}

\begin{table}[!t]
\centering
\caption{The MCC values of the methods.}
\vspace{-0.3cm}
\label{tab:mcc}
\Huge  % 整体字体缩小
\renewcommand{\arraystretch}{1}  % 行间距加宽，提升可读性

% 第一行：Qwen2.5-Coder + CodeGemma
\begin{subtable}[t]{0.48\linewidth}
\centering
\caption{Qwen2.5-Coder}
\label{tab:mcc_a}
\resizebox{1.05\linewidth}{!}{  % 自动适配宽度
\begin{tabular}{c|cccccccccccc}
        \hline
        \textbf{Bench} &\textbf{Go} &\textbf{DL} &\textbf{CI}  &\textbf{RF} &\textbf{LR} &\textbf{CNN} &\textbf{LSTM} &\textbf{Trans.} &\textbf{Ours} \\
        \hline
        HP & 0.17  & 0.10  & 0.24  & 0.59 & 0.31  & 0.57  & 0.15  & 0.40  & \textbf{0.77}  \\
        HJ & 0.53  & \textbf{0.87}  & 0.45  & 0.88 & 0.33  & 0.44  & 0.26  & 0.39  & 0.80  \\
        MP & 0.30  & 0.34  & 0.27  & 0.52  & 0.29  & 0.34  & 0.04  & 0.25  & \textbf{0.71}  \\
        MJ & 0.42  & 0.65  & 0.24  & 0.53  & 0.23  & 0.32  & 0.10  & 0.28  & \textbf{0.75}  \\
        NP & 0.54  & 0.00  & 0.26  & 0.23  & 0.04  & \textbf{0.76}  & 0.75  & 0.60  & 0.68  \\
        NJ & \textbf{0.93}  & 0.36  & 0.00  & 0.62 & 0.00  & 0.77  & 0.72  & 0.64  & 0.91  \\
        EB & 0.40  & 0.84  & 0.27  & 0.70  & 0.09  & 0.54  & \textbf{0.82}  & 0.42  & 0.69  \\
        CE & 0.78  & 0.45  & 0.36  & 0.53  & 0.35  & 0.64  & \textbf{0.97}  & 0.44  & 0.67  \\
        \hline
        AVG & 0.51  & 0.45  & 0.26  & 0.58  & 0.20  & 0.55  & 0.48  & 0.43 & \textbf{0.75} \\ \hline
        W/D/L & 6/0/2 & 6/0/2 & 8/0/0 & 6/0/2 & 8/0/0 & 7/0/1 & 5/0/3 & 8/0/0 & \\
        \hline
        p-value & \textbf{0.04} & 0.25 & \textbf{0.01} & \textbf{0.04} & \textbf{0.02} & \textbf{0.03} & 0.05 & \textbf{0.03} &  \\
        \hline
        $\delta$ & \textbf{0.59} & \textbf{0.53} & \textbf{1.00} & \textbf{0.78} & \textbf{1.00} & \textbf{0.63} & \textbf{0.59} & \textbf{0.97} & \\
        \hline
        \end{tabular}
}
\end{subtable}
\hfill  % 两列间距自动填充
\begin{subtable}[t]{0.48\linewidth}
\centering
\caption{CodeGemma}
\label{tab:mcc_b}
\resizebox{1.05\linewidth}{!}{
\begin{tabular}{c|cccccccccccc}
        \hline
        \textbf{Bench} &\textbf{Go} &\textbf{DL} &\textbf{CI}  &\textbf{RF} &\textbf{LR} &\textbf{CNN} &\textbf{LSTM} &\textbf{Trans.} &\textbf{Ours} \\
        \hline
        HP & 0.17 & 0.10 & 0.10 & 0.22 & 0.03 & 0.15 & 0.40 & 0.57 & \textbf{0.79}\\
        HJ & 0.53 & \textbf{0.87} & 0.42 & 0.86 & 0.38 & 0.26 & 0.39 & 0.44 & 0.78\\
        MP & 0.30 & 0.34 & 0.36 & 0.40 & 0.23 & 0.04 & 0.25 & 0.34 & \textbf{0.76}\\
        MJ & 0.42 & 0.65 & 0.23 & 0.59 & 0.20 & 0.10 & 0.28 & 0.32 & \textbf{0.69}\\
        NP & 0.54 & 0.00 & 0.27 & 0.17 & 0.09 & 0.75 & 0.60 & 0.76 & \textbf{0.91}\\
        NJ & 0.93 & 0.36 & 0.15 & 0.46 & 0.04 & 0.72 & 0.64 & 0.77 & \textbf{1.00}\\
        EB & 0.40 & \textbf{0.84} & 0.21 & 0.81 & 0.12 & 0.82 & 0.42 & 0.54 & 0.78\\
        CE & 0.78 & 0.45 & 0.37 & 0.80 & 0.15 & \textbf{0.97} & 0.44 & 0.64 & 0.75\\
        \hline
        AVG & 0.51 & 0.45 & 0.26 & 0.54 & 0.16 & 0.48 & 0.43 & 0.55 & \textbf{0.81}\\
        \hline
        W/D/L & 7/0/1 & 6/0/2 & 8/0/0 & 5/0/3 & 8/0/0 & 6/0/2 & 8/0/0 & 8/0/0 & \\
        \hline
        p-value & \textbf{0.04} & 0.05 & \textbf{0.01}  & 0.11 & \textbf{0.02} & \textbf{0.03} & 0.15 & \textbf{0.03} & \\
        \hline
        $\delta$ & \textbf{0.56} & \textbf{0.56} & \textbf{1.00} & \textbf{0.34} & \textbf{1.00} & \textbf{0.69} & \textbf{0.28} & \textbf{1.00} & \\
        \hline
        \end{tabular}
}
\end{subtable}

\vspace{1pt}  % 两行间距

% 第二行：DeepSeek-Coder + Phi-2（假设你补充对应真实数据）
\begin{subtable}[t]{0.48\linewidth}
\centering
\caption{DeepSeek-Coder}
\label{tab:mcc_c}
\resizebox{1.05\linewidth}{!}{
\begin{tabular}{c|cccccccccccc}
        \hline
        \textbf{Bench} &\textbf{Go} &\textbf{DL} &\textbf{CI} &\textbf{RF} &\textbf{LR} &\textbf{CNN} &\textbf{LSTM} &\textbf{Trans.} &\textbf{Ours} \\
        \hline
        HP & 0.23  & 0.53  & 0.43  & 0.53  & 0.30  & 0.56  & 0.18  & 0.33  & \textbf{0.70}  \\
        HJ & 0.53  & \textbf{0.99}  & 0.25  & 0.96 & 0.18  & 0.47  & 0.24  & 0.36  & 0.79  \\
        MP & 0.29  & 0.74  & 0.34  & 0.69  & 0.35  & 0.37  & 0.05  & 0.26  & \textbf{0.71}  \\
        MJ & 0.43  & 0.66  & 0.22  & 0.63  & 0.20  & 0.34  & 0.16  & 0.26  & \textbf{0.75}  \\
        NP & \textbf{0.57}  & 0.00  & 0.25  & 0.16  & 0.04  & 0.72  & 0.69  & 0.54  & 0.56  \\
        NJ & \textbf{0.92}  & 0.16  & 0.00  & 0.24  & 0.00  & 0.77  & 0.72  & 0.64  & 0.91  \\
        EB & 0.44  & 0.55  & 0.19  & 0.44  & 0.11  & 0.51  & \textbf{0.73}  & 0.39  & 0.64  \\
        CE & 0.77  & 0.35  & 0.15  & 0.57  & 0.14  & 0.64  & \textbf{0.97}  & 0.44 & 0.80  \\
        \hline
        AVG & 0.52  & 0.50  & 0.23  & 0.53  & 0.16  & 0.55  & 0.47  & 0.40  & \textbf{0.73} \\ \hline
        W/D/L & 6/0/2 & 6/0/2 & 8/0/0 & 7/0/1 & 8/0/0 & 7/0/1 & 5/0/3 & 8/0/0 & \\
        \hline
        p-value & \textbf{0.04} & 0.11 & \textbf{0.01} & 0.05 & \textbf{0.02} & 0.05 & 0.11 & \textbf{0.03} & \\
        \hline
        $\delta$ & \textbf{0.56} & \textbf{0.56} & \textbf{1.00} & \textbf{0.63} & \textbf{1.00} & \textbf{0.66} & \textbf{0.44} & \textbf{0.94} & \\
        \hline
        \end{tabular}
}
\end{subtable}
\hfill
\begin{subtable}[t]{0.48\linewidth}
\centering
\caption{Phi-2}
\label{tab:mcc_d}
\resizebox{1.05\linewidth}{!}{
\begin{tabular}{c|cccccccccccc}
        \hline
        \textbf{Bench} &\textbf{Go} &\textbf{DL} &\textbf{CI} &\textbf{RF} &\textbf{LR} &\textbf{CNN} &\textbf{LSTM} &\textbf{Trans.} &\textbf{Ours} \\
        \hline
        HP & 0.13  & 0.19  & 0.00  & 0.24  & 0.27  & 0.23  & 0.02  & 0.06  & \textbf{0.79}  \\
        HJ & 0.15  & 0.32  & 0.32  & 0.00 & 0.39  & 0.23  & 0.10  & 0.16  & \textbf{0.77}  \\
        MP & 0.15  & 0.24  & 0.25  & 0.47  & 0.48  & 0.18  & 0.02  & 0.05  & \textbf{0.61}  \\
        MJ & 0.17  & 0.47  & 0.05  & \textbf{0.64}  & 0.49  & 0.15  & 0.06  & 0.00  & 0.63  \\
        NP & 0.20  & 0.23  & 0.08  & 0.00  & 0.42  & 0.19  & 0.27  & 0.10  & \textbf{0.59}  \\
        NJ & 0.16 & 0.30  & 0.00  & 0.38  & 0.46  & 0.27  & 0.49  & 0.00  & \textbf{0.54}  \\
        EB & 0.17  & 0.15  & 0.14  & \textbf{0.64}  & 0.42  & 0.21  & 0.24  & 0.04  & 0.53  \\
        CE & 0.47  & 0.42  & 0.19  & 0.37  & 0.44  & 0.34  & \textbf{0.74}  & 0.03 & 0.55  \\
        \hline
        AVG & 0.20  & 0.29  & 0.11  & 0.34  & 0.42  & 0.22  & 0.23  & 0.05  & \textbf{0.63} \\ \hline
        W/D/L & 8/0/0 & 8/0/0 & 8/0/0 & 6/0/2 & 8/0/0 & 8/0/0 & 7/0/1 & 8/0/0 & \\
        \hline
        p-value & \textbf{0.00} & \textbf{0.00} & \textbf{0.01} & \textbf{0.04} & \textbf{0.01} & \textbf{0.00} & \textbf{0.01} & \textbf{0.00} & \\
        \hline
        $\delta$ & \textbf{0.43} & \textbf{0.34} & \textbf{0.52} & \textbf{0.29} & \textbf{0.21} & \textbf{0.40} & \textbf{0.40} & \textbf{0.58} & \\
        \hline
        \end{tabular}
}
\end{subtable}
\vspace{-10pt}
\end{table}

\begin{table}[!t]
\centering
\caption{The AUC values of the methods.}
\vspace{-0.3cm}
\label{tab:auc}
\Huge  % 整体字体缩小
\renewcommand{\arraystretch}{1}  % 行间距加宽，提升可读性

% 第一行：Qwen2.5-Coder + CodeGemma
\begin{subtable}[t]{0.48\linewidth}
\centering
\caption{Qwen2.5-Coder}
\label{tab:auc_a}
\resizebox{1.05\linewidth}{!}{  % 自动适配宽度
\begin{tabular}{c|cccccccccccc}
        \hline
        \textbf{Bench} &\textbf{Go} &\textbf{DL} &\textbf{CI} &\textbf{RF} &\textbf{LR} &\textbf{CNN} &\textbf{LSTM} &\textbf{Trans.} &\textbf{Ours} \\
        \hline
        HP & 0.68  & 0.78  & 0.75  & 0.90  & 0.48  & 0.85  & 0.61  & 0.74  & \textbf{0.95}  \\
        HJ & 0.88  & \textbf{0.99}  & 0.74  & 0.95  & 0.72  & 0.79  & 0.64  & 0.73  & 0.94  \\
        MP & 0.74  & 0.78  & 0.73  & 0.85  & 0.64  & 0.74  & 0.52  & 0.65  & \textbf{0.92}  \\
        MJ & 0.77  & 0.75  & 0.63  & 0.86  & 0.60  & 0.71  & 0.55  & 0.69  & \textbf{0.93}  \\
        NP & 0.89  & 0.50  & 0.63  & 0.62  & 0.55  & 0.94  & 0.94  & 0.87  & \textbf{1.00} \\
        NJ & 0.98  & 0.69  & 0.35  & 0.75  & 0.55  & 0.98  & 0.97  & 0.90  & \textbf{1.00}  \\
        EB & 0.80  & 0.89  & 0.66  & 0.92  & 0.64  & 0.87  & \textbf{0.95}  & 0.79  & 0.93  \\
        CE & 0.97  & 0.81  & 0.70  & 0.91  & 0.63  & 0.93  & \textbf{1.00}  & 0.80  & 0.91  \\
        AVG  & 0.84  & 0.77  & 0.65  & 0.85  & 0.60  & 0.85  & 0.77  & 0.77  & \textbf{0.95} \\ \hline
        W/D/L & 7/0/1 & 7/0/1 & 8/0/0 & 6/0/2 & 8/0/0 & 7/0/1 & 6/0/2 & 8/0/0 & \\
        \hline
        p-value & \textbf{0.03} & \textbf{0.04} & \textbf{0.01} 
        & 0.08 & \textbf{0.02} & \textbf{0.03} & 0.08 & \textbf{0.02} &  \\
        \hline
        $\delta$ & \textbf{0.63} & \textbf{0.81} & \textbf{1.00} & \textbf{0.75} & \textbf{1.00} & \textbf{0.63} & \textbf{0.28} & \textbf{1.00} & \\
        \hline
        \end{tabular}
}
\end{subtable}
\hfill  % 两列间距自动填充
\begin{subtable}[t]{0.48\linewidth}
\centering
\caption{CodeGemma}
\label{tab:auc_b}
\resizebox{1.05\linewidth}{!}{
\begin{tabular}{c|cccccccccccc}
        \hline
        \textbf{Bench} &\textbf{Go} &\textbf{DL} &\textbf{CI} &\textbf{RF} &\textbf{LR} &\textbf{CNN} &\textbf{LSTM} &\textbf{Trans.} &\textbf{Ours} \\
        \hline
        HP & 0.63  & 0.51  & 0.72  & 0.81  & 0.73  & 0.85  & 0.57  & 0.76  & \textbf{0.95}  \\
        HJ & 0.83  & 0.93  & 0.76  & \textbf{0.98}  & 0.76  & 0.81  & 0.64  & 0.75  & \textbf{0.98}  \\
        MP & 0.71  & 0.62  & 0.67  & 0.84  & 0.68  & 0.71  & 0.49  & 0.68  & \textbf{0.91}  \\
        MJ & 0.78  & 0.82  & 0.65  & 0.88  & 0.65  & 0.73  & 0.54  & 0.69  & \textbf{0.94}  \\
        NP & 0.86  & 0.50  & 0.61  & 0.62  & 0.62  & 0.93  & 0.95  & 0.87  & \textbf{0.97} \\
        NJ & \textbf{0.99}  & 0.61  & 0.35  & 0.68  & 0.35  & 0.98  & 0.97  & 0.90  & \textbf{0.99}  \\
        EB & 0.77  & 0.92  & 0.61  & 0.96  & 0.61  & 0.87  & \textbf{0.97}  & 0.78  & 0.92  \\
        CE & 0.95  & 0.67  & 0.76  & 0.98  & 0.81  & 0.93  & \textbf{1.00}  & 0.80  & 0.89  \\
        \hline
        AVG & 0.81  & 0.70  & 0.64  & 0.84  & 0.65  & 0.85  & 0.77  & 0.78  & \textbf{0.94} \\
        \hline
        W/D/L & 7/0/1 & 7/1/0 & 8/0/0 & 5/1/2 & 8/0/0 & 7/0/1 & 6/0/2 & 8/0/0 & \\
        \hline
        p-value & \textbf{0.03} & \textbf{0.00} & \textbf{0.01} & 0.19 & \textbf{0.03} & \textbf{0.04} & 0.07 & \textbf{0.03} & \\
        \hline
        $\delta$ & \textbf{0.65} & \textbf{0.88} & \textbf{1.00} & \textbf{0.41} & \textbf{1.00} & \textbf{0.68} & \textbf{0.28} & \textbf{0.98} & \\
        \hline
        \end{tabular}
}
\end{subtable}

\vspace{1pt}  % 两行间距

% 第二行：DeepSeek-Coder + Phi-2（假设你补充对应真实数据）
\begin{subtable}[t]{0.48\linewidth}
\centering
\caption{DeepSeek-Coder}
\label{tab:auc_c}
\resizebox{1.05\linewidth}{!}{
\begin{tabular}{c|ccccccccc}
\toprule
\textbf{Bench} &\textbf{Go} &\textbf{DL} &\textbf{CI}  &\textbf{RF} &\textbf{LR} &\textbf{CNN} &\textbf{LSTM} &\textbf{Trans.} &\textbf{Ours} \\
\hline
% 这里替换为 DeepSeek-Coder 的真实数据，示例如下（需与你的实验一致）：
HP & 0.68  & 0.72  & 0.75 & 0.81 & 0.72  & 0.85  & 0.60  & 0.74  & \textbf{0.94}  \\
HJ & 0.85  & \textbf{1.00}  & 0.67  & 0.97 & 0.67  & 0.82  & 0.64  & 0.77  & 0.96  \\
MP & 0.70  & 0.87  & 0.70  & 0.84 & 0.70  & 0.74  & 0.53  & 0.67  & \textbf{0.92}  \\
MJ & 0.79  & 0.81  & 0.66  & 0.88 & 0.65  & 0.74  & 0.57  & 0.69  & \textbf{0.95}  \\
NP & 0.88  & 0.50  & 0.60  & 0.62 & 0.61  & 0.94  & 0.94  & 0.85  & \textbf{0.94}  \\
NJ & 0.99  & 0.52  & 0.35  & 0.68 & 0.35  & 0.98  & 0.97  & 0.90  & \textbf{1.00}  \\
EB & 0.80  & 0.76  & 0.58  & \textbf{0.96} & 0.58  & 0.85  & 0.95  & 0.78  & 0.93  \\
CE & 0.95  & 0.65  & 0.67  & 0.98 & 0.65  & 0.93  & \textbf{1.00}  & 0.80  & 0.94  \\
AVG & 0.83  & 0.73  & 0.62  & 0.85  & 0.62  & 0.86  & 0.77  & 0.77 & \textbf{0.95} \\ 
\hline
W/D/L & 7/0/1 & 7/0/1 & 8/0/0 & 5/0/3 & 8/0/0 & 7/1/0 & 5/1/2 & 8/0/0 & \\
\hline
p-value & \textbf{0.03} & \textbf{0.04} & \textbf{0.01} & 0.19 & \textbf{0.02} & \textbf{0.04} & 0.15 & \textbf{0.03} & \\
\hline
$\delta$ & \textbf{0.59} & \textbf{0.78} & \textbf{1.00} & \textbf{0.41} & \textbf{1.00} & \textbf{0.59} & \textbf{0.28} & \textbf{1.00} & \\
\bottomrule
\end{tabular}
}
\end{subtable}
\hfill
\begin{subtable}[t]{0.48\linewidth}
\centering
\caption{Phi-2}
\label{tab:auc_d}
\resizebox{1.05\linewidth}{!}{
\begin{tabular}{c|ccccccccc}
\toprule
\textbf{Bench} &\textbf{Go} &\textbf{DL} &\textbf{CI}  &\textbf{RF} &\textbf{LR} &\textbf{CNN} &\textbf{LSTM} &\textbf{Trans.} &\textbf{Ours} \\
\hline
HP & 0.58  & 0.13  & 0.59  & 0.68  & 0.69  & 0.70  & 0.50  & 0.59  & \textbf{0.98}  \\
        HJ & 0.61  & 0.31  & 0.72  & 0.50  & 0.69  & 0.63  & 0.53  & 0.60  & \textbf{0.97}  \\
        MP & 0.60  & 0.20  & 0.65  & 0.83  & 0.67  & 0.63  & 0.51  & 0.53  & \textbf{0.89} \\
        MJ & 0.61  & 0.63  & 0.58  & \textbf{0.87}  & 0.74  & 0.61  & 0.51  & 0.52  & \textbf{0.87}  \\
        NP & 0.62  & 0.35  & 0.49  & 0.50  & 0.675  & 0.66  & 0.71  & 0.57  & \textbf{0.82}  \\ 
        NJ & 0.62  & 0.28  & 0.42  & \textbf{0.96}  & 0.44  & 0.74  & 0.83  & 0.53  & 0.81  \\
        EB & 0.63  & 0.08  & 0.61  & \textbf{0.95}  & 0.79  & 0.64  & 0.71  & 0.55  & 0.89  \\
        CE & 0.83  & 0.46  & 0.67  & 0.76  & 0.62  & 0.69  & \textbf{0.94}  & 0.47  & 0.88  \\
        \hline
        AVG & 0.64  & 0.31  & 0.59  & 0.76  & 0.67  & 0.66  & 0.66  & 0.54  & \textbf{0.89} \\ \hline
        W/D/L & 8/0/0 & 8/0/0 & 8/0/0 & 5/1/2 & 8/0/0 & 8/0/0 & 6/0/2 & 8/0/0 & \\
        \hline
        p-value & \textbf{0.01} & \textbf{0.00} & \textbf{0.00} & 0.12 & \textbf{0.00} & \textbf{0.01} & \textbf{0.01} & \textbf{0.01} & \\
        \hline
        $\delta$ & \textbf{0.25} & \textbf{0.58} & \textbf{0.29} & \textbf{0.13} & \textbf{0.22} & \textbf{0.22} & \textbf{0.23} & \textbf{0.34} & \\
\bottomrule
\end{tabular}
}
\end{subtable}
\vspace{-10pt}
\end{table}

\begin{tcolorbox}[
    colback=gray!5,      % Background color
    colframe=gray!50,    % Border color
    coltitle=black,      % Title text color
    boxsep=0.01pt,       % Space between box and text
    left=3pt,             % Left padding
    right=3pt,            % Right padding
    rounded corners,      % Rounded corners for the box
    fonttitle=\bfseries        % Bold title font
]
\textbf{Answer to RQ1:}  CGMIA significantly outperforms other MIA methods for data leakage detection in code generation benchmarks in the vast majority of cases.
\end{tcolorbox}

\subsection{\textbf{RQ2: How do the proposed expert features influence the performance of CGMIA?}}

\begin{figure*}[!t]
\centering
\includegraphics[width=0.95\textwidth]{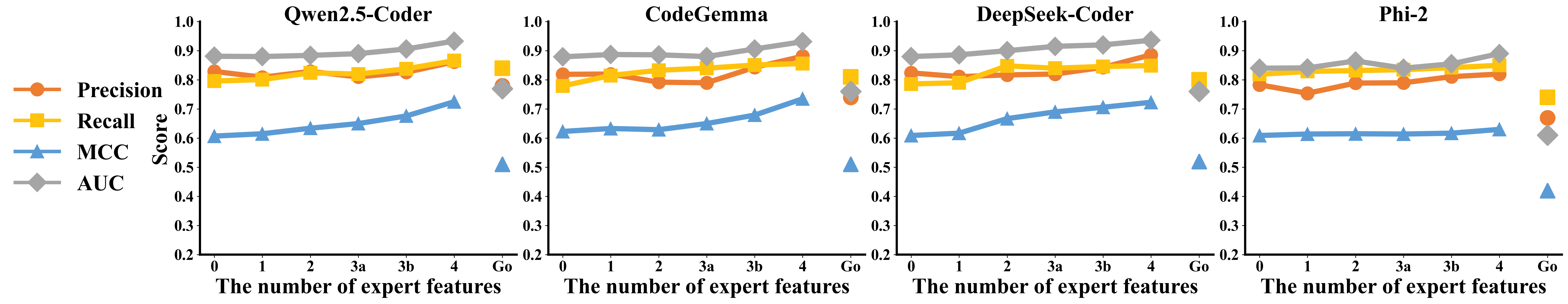}
\caption{The performance of CGMIA under varying numbers of expert features.}
\label{rq2_1}
\end{figure*}

\begin{figure*}[!t]
\centering
\includegraphics[width=0.95\textwidth]{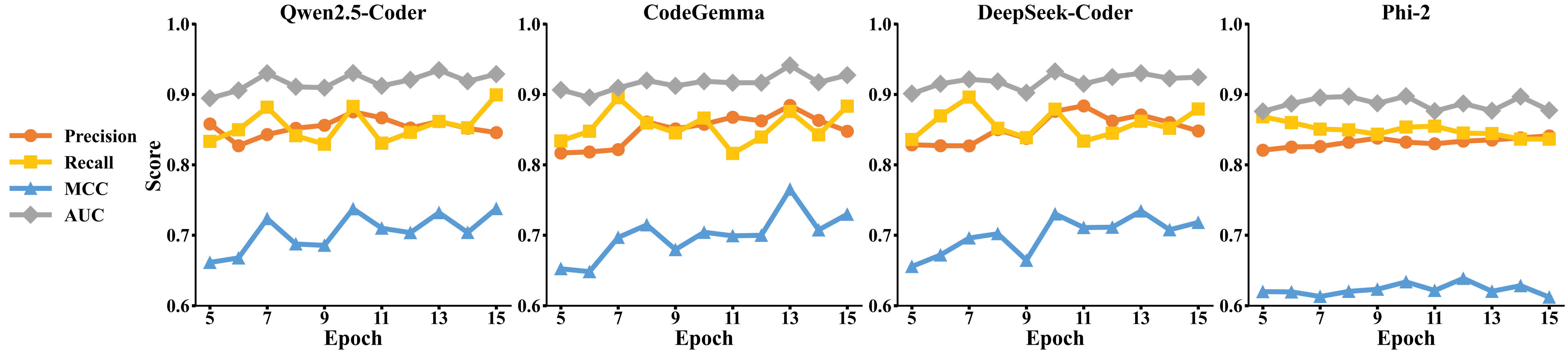}
\caption{The impact of different fine-tuning epochs of target models on the performance of CGMIA.}
\label{rq2_2}
\end{figure*}

\begin{figure}[!t]
    \centering
    
        % ------------------- 第1行：2个子图横向排列 -------------------
    \begin{subfigure}[b]{0.49\linewidth} % 每个子图占页面48%宽度（2个合计96%，留4%间距）
        \centering
        % 子图内部宽度占满subfigure，保证图片比例一致
        \includegraphics[width=\linewidth]{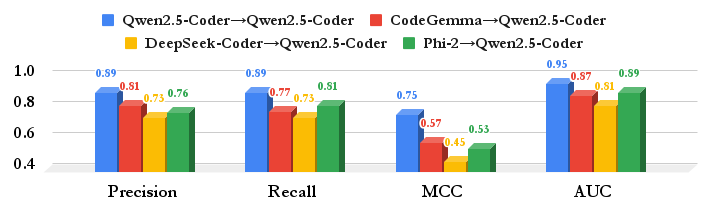}
        \caption{Using Qwen2.5-Coder as the target model}
        \label{fig:qwen}
    \end{subfigure}
    \hfill % 自动填充两子图间空白，确保第1行左右对齐
    \begin{subfigure}[b]{0.49\linewidth} % 与左侧子图宽度严格一致（保证对齐）
        \centering
        \includegraphics[width=\linewidth]{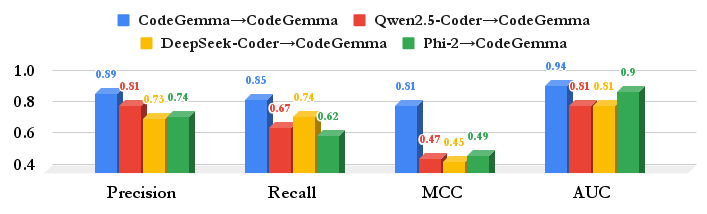}
        \caption{Using CodeGemma as the target model}
        \label{fig:codegemma}
    \end{subfigure}
    
    % ------------------- 第2行：2个子图横向排列（与第1行对齐） ------------------- 
    \begin{subfigure}[b]{0.49\linewidth} % 宽度与第1行保持一致（0.48\linewidth）
        \centering
        \includegraphics[width=\linewidth]{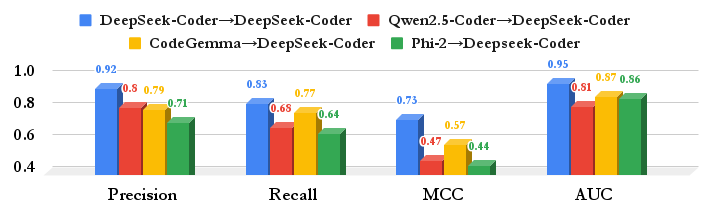}
        \caption{Using DeepSeek-Coder as the target model}
        \label{fig:deepseek} % 注意：原代码中2个deepseek用了同一label，此处已修正为唯一label
    \end{subfigure}
    \hfill % 与第1行逻辑一致，保证第2行左右对齐
    \begin{subfigure}[b]{0.49\linewidth} % 宽度统一，确保2×2布局整齐
        \centering
        \includegraphics[width=\linewidth]{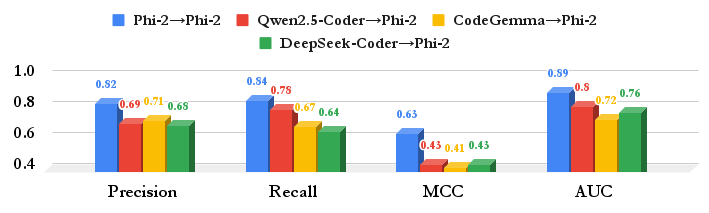} % 修正图片路径（原重复用了deepseek.png）
        \caption{Using Phi-2 as the target model}
        \label{fig:deepseek} % 新增唯一label，便于后文引用
    \end{subfigure}
    
    % 整体图标题（调整与上方子图的间距）
    \caption{The impact of different shadow models on the data leakage detection performance of CGMIA.}
    \label{fig:target_change}
\end{figure}

We propose the four expert features, i.e., CodeBLEU, Edit Distance, Perplexity, and Test Pass Rate, and integrate them with semantic features extracted by CodeBERT for membership inference classifier training. Given the heterogeneous nature of these features and their varying impact on classification performance, we systematically analyze all 16 ($=C_4^1+C_4^2+C_4^3+C_4^4+1$) possible feature combinations (including the case with no expert features) to assess their effectiveness. We categorize these combinations into five levels based on the number of expert features included: Level-0 includes no expert features (semantic features only); Level-1 includes combinations with one feature; Level-2 with two; Level-3 with three; and Level-4 with all four expert features. For each level, we report the best-performing combination based on the average results across eight code generation benchmarks in Figure \ref{rq2_1}. These combinations are as follows: Level-1: \textit{Perplexity}; Level-2: \textit{Perplexity} + \textit{Test Pass Rate}; Level-3b: \textit{Perplexity} + \textit{Test Pass Rate} + \textit{CodeBLEU}; Level-4: \textit{Perplexity} + \textit{Test Pass Rate} + \textit{CodeBLEU} + \textit{Edit Distance}. 
While most code generation benchmarks provide test cases for calculating Test Pass Rate, a small minority do not include such test cases \cite{feng2024complexcodeeval}. Thus, for Level-3 (which uses three expert features), we additionally define a variant—Level-3a, where the Test Pass Rate feature is excluded. This variant is designed to evaluate the CGMIA’s robustness when Test Pass Rate is unavailable due to missing benchmark test cases. By contrast, Level-3b (\textit{Perplexity} + \textit{Test Pass Rate} + \textit{CodeBLEU}) represents the best-performing combination among all three-feature subsets of expert features. 
Additionally, the performance of Gotcha is included in Figure \ref{rq2_1}, displayed as scatter points on the far right of each subplot. This setup enables a clear comparison between Gotcha and Level-0, where CGMIA relies solely on semantic features. 

Experimental results show that as the number of expert features increases, our CGMIA method achieves progressively better performance across all four metrics. Using all expert features yields the highest scores across all models, with a single exception where the Recall on the DeepSeek-Coder model is marginally higher at Level-2. These findings demonstrate that each expert feature provides complementary benefits and that combining them enhances the effectiveness of our CGMIA method. Furthermore, our results for Level-3a and Level-4 indicate that while Level-4 outperforms Level-3a, the performance difference between them is relatively small, with the gap not exceeding 0.05 across all metrics. This indicates that CGMIA remains effective even when Test Pass Rate is unavailable, thereby demonstrating the robustness of our feature design in the absence of test case-dependent metrics. Additionally, comparative results between Level-0 and Gotcha reveal that Level-0 consistently outperforms Gotcha across all four metrics. This superiority can be attributed to key limitations in Gotcha’s design: Gotcha underperforms in both feature extraction and model construction, as it simply feeds extracted features directly into a binary classifier without the classifier participating in the network training process.

\begin{tcolorbox}[
    colback=gray!5,      % Background color
    colframe=gray!50,    % Border color
    coltitle=black,      % Title text color
    boxsep=0.01pt,       % Space between box and text
    left=3pt,             % Left padding
    right=3pt,            % Right padding
    rounded corners,      % Rounded corners for the box
    fonttitle=\bfseries        % Bold title font
]
\textbf{Answer to RQ2:}  Each expert feature contributes to enhancing CGMIA's performance. 
\end{tcolorbox}

\subsection{\textbf{RQ3: What are the factors affecting the data leakage detection of CGMIA?}}

In this RQ, we examine the robustness of CGMIA from two perspectives. First, we vary the number of fine-tuning epochs of the target model to evaluate whether CGMIA is sensitive to the leakage simulation strength. Second, we study shadow-target model mismatch, where the shadow model and the target model come from different model families, to evaluate whether CGMIA remains effective in a more realistic black-box setting.

\textbf{The fine-tuning epochs of target models.} 
Prior research~\cite{yeom2018privacy} notes that membership leakage risk may rise with more epochs. In RQ1, we fine-tune each target model for 10 epochs. To examine whether CGMIA depends on this specific setting, we fine-tune target models for 5, 6, 7, ..., and 15 epochs. Each point on the line chart in Figure~\ref{rq2_2} represents the average value of the CGMIA method across these eight code generation benchmarks. The results indicate that the four metrics remain relatively stable with only minor fluctuations. We use the Wilcoxon signed-rank test to check for significant differences in CGMIA's performance on 8 benchmarks between 10 epochs and other counts, finding no significant difference. These results indicate that CGMIA is not overly sensitive to the exact number of fine-tuning epochs used in our leakage simulation. In other words, CGMIA can still identify membership signals even when the target model is fine-tuned for fewer epochs.

\textbf{The shadow models.} 
In RQ1, we use the same LLMs for both the target and shadow models. For example, when Qwen2.5-Coder is used as the target model, Qwen2.5-Coder is also used as the shadow model. However, in practical black-box scenarios, the attacker usually does not know the exact architecture of the target LLM. Therefore, this same-model setting represents a relatively favorable evaluation setting, and it is important to further examine CGMIA under shadow-target architectural mismatch. For this RQ, we analyze the impact on our CGMIA method by using shadow models with different architectures from the target model. For example, in Figure~\ref{fig:qwen}, Qwen2.5-Coder serves as the target model, while Qwen2.5-Coder, CodeGemma, DeepSeek-Coder, and Phi-2 are used as shadow models. Each bar of Figure~\ref{fig:target_change} shows the average value of the CGMIA method across eight code generation benchmarks. An asterisk (*) after the average indicates a significant difference between the results of using an LLM with a different architecture as the shadow model and those of using an LLM with the same architecture as the shadow model across these eight benchmarks.

Figure~\ref{fig:target_change} shows that CGMIA exhibits a moderate performance decline when using shadow models with different architectures. Specifically, MCC decreases statistically significantly, while Precision, Recall, and AUC only show non-significant drops in most cases. Even so, CGMIA remains highly effective, with key metrics maintaining strong ranges: Precision 0.73--0.81, Recall 0.68--0.77, AUC 0.81--0.87, and MCC 0.41--0.57, indicating that CGMIA still provides useful detection results under shadow-target mismatch. Under the controlled evaluation protocol, the Precision values indicate that a considerable proportion of samples predicted as leaked are correctly identified, while the Recall values show that CGMIA can still identify many member samples. The AUC values further indicate that CGMIA maintains reasonable overall discrimination ability even when the shadow model differs from the target model. This also suggests that the manually designed expert features and CodeBERT-based semantic features can still provide robust membership signals for CGMIA, enabling it to maintain good detection performance even when the target and shadow models have different architectures.

Additionally, Table~\ref{tab:rfandgotcha} presents the average values of the top two performing baselines, random forest and Gotcha, across eight benchmarks when the target model and shadow model differ. In Table~\ref{tab:rfandgotcha}, C denotes CodeGemma, Q denotes Qwen2.5-Coder, D denotes DeepSeek-Coder, and P denotes Phi-2. For example, cell ``0.77/0.65'' indicates that using CodeGemma as the shadow model and Qwen2.5-Coder as the target model, random forest and Gotcha achieve average precision values of 0.77 and 0.65 on the eight benchmarks, respectively. Asterisks (*) in Table~\ref{tab:rfandgotcha} denote cases where our CGMIA method significantly outperforms random forest and Gotcha. Comparing the results in Figure~\ref{fig:target_change} and Table~\ref{tab:rfandgotcha}, we observe that our CGMIA method still achieves higher average values than random forest and Gotcha across all eight benchmarks. In some cases, CGMIA significantly outperforms random forest and Gotcha. For instance, when using CodeGemma as the shadow model and Qwen2.5-Coder as the target model, CGMIA achieves an AUC value of 0.87, which is significantly higher than Gotcha's AUC of 0.70. This comparison shows that the advantage of CGMIA is still observed in the mismatched shadow-model setting. Random forest relies on expert features, while Gotcha mainly relies on semantic embeddings. CGMIA combines both types of features, and the results are consistent with the expectation that integrating expert and semantic information can improve leakage detection performance under architectural mismatch.

\begin{tcolorbox}[
    colback=gray!5,      % Background color
    colframe=gray!50,    % Border color
    coltitle=black,      % Title text color
    boxsep=0.01pt,       % Space between box and text
    left=3pt,             % Left padding
    right=3pt,            % Right padding
    rounded corners,      % Rounded corners for the box
    fonttitle=\bfseries        % Bold title font
]
\textbf{Answer to RQ3:}  The number of fine-tuning epochs in target models has little impact on data leakage detection performance. While the choice of shadow model impacts CGMIA's performance, CGMIA still retains strong performance, consistently showing practical reliability and usability. 
\end{tcolorbox}

\subsection{RQ4: Can ensemble learning mitigate the performance degradation of CGMIA caused by architectural mismatch between shadow and target models?}

Figure \ref{fig:target_change} in RQ3 reveals that selecting different shadow models for a given target model leads to varying CGMIA performance outcomes. For instance, when Qwen 2.5-Coder is the target model, DeepSeek-Coder (as the shadow model) yields the lowest MCC and AUC values, while CodeGemma achieves the highest; when DeepSeek-Coder is the target, Phi-2 results in the lowest MCC and AUC, whereas CodeGemma delivers the highest; for CodeGemma as the target, Qwen 2.5-Coder produces the lowest MCC and AUC, while Phi-2 achieves the highest; and when Phi-2 is the target, CodeGemma yields the lowest MCC and AUC, with Qwen 2.5-Coder achieving the highest. Overall, no consistent pattern emerges for choosing the optimal or suboptimal shadow models. To address this challenge, we propose an ensemble learning strategy that integrates three of the four models as shadow models, with the remaining as the target. Each shadow model outputs a binary prediction (1 for leakage, 0 for no leakage). We assign weights to the shadow models based on their MCC scores on validation sets, normalizing these scores so their sum equals 1. The final decision is made by weighted voting: each prediction is multiplied by its weight, and the sum determines the outcome. If the total score exceeds 1, the sample is classified as leaked; otherwise, it is non-leaked. 

Table \ref{tab:ensemble} presents the average performance of the ensemble learning strategy across the eight benchmarks. For instance, the notation “C/D/P→Q” denotes an ensemble of three shadow models (CodeGemma, DeepSeek-Coder, Phi-2) employed to detect membership leakage in the target model Qwen 2.5-Coder. Specifically, the first column (labeled “D→Q”) shows the performance when DeepSeek-Coder is used alone as the shadow model—this combination yields the worst MCC and AUC results for Qwen 2.5-Coder detection. Conversely, the second column (labeled “C→Q”) demonstrates that using CodeGemma alone achieves the best MCC and AUC for the same target model. Experimental results demonstrate that the ensemble learning strategy outperforms the worst-performing single shadow model in nearly all cases, except when CodeGemma is the target model, where the ensemble’s Recall value (0.72) is slightly lower than that of Qwen 2.5-Coder alone as the shadow model. Moreover, among the 16 total cases (4 target models × 4 metrics), the ensemble even surpasses the best-performing single shadow model in 10 cases (62.5\%). 
These results show that the ensemble strategy improves robustness under unknown target architectures. Since the best single shadow model cannot be identified in advance in the black-box setting, the main value of the ensemble is to reduce the worst-case risk of selecting a poorly matched shadow model. Although the ensemble does not always outperform the best single shadow model, it consistently performs better than the worst single shadow model in most cases. This makes the ensemble strategy a more reliable choice when the attacker has no architectural knowledge of the target model.
% Therefore, when the architecture of the target model is unknown and the choice of shadow model is uncertain, adopting an ensemble learning strategy provides a more reliable and stable alternative. Although the ensemble may not always outperform the optimal single shadow model, it consistently ensures performance that is substantially better than the worst-case single shadow model scenario. 

\begin{tcolorbox}[
    colback=gray!5,      % Background color
    colframe=gray!50,    % Border color
    coltitle=black,      % Title text color
    boxsep=0.01pt,       % Space between box and text
    left=3pt,             % Left padding
    right=3pt,            % Right padding
    rounded corners,      % Rounded corners for the box
    fonttitle=\bfseries        % Bold title font
]
\textbf{Answer to RQ4:}  Our proposed ensemble strategy mitigates CGMIA’s performance degradation from shadow-target architectural mismatch, and though not always outperforming the optimal single shadow model, it consistently outperforms the worst single shadow model—serving as a more reliable alternative. 
\end{tcolorbox}

\begin{table*}[t]
\centering
\caption{The performance of CGMIA and baselines on StarCoder.}
\label{tab:starcoder_all_baselines}
\scriptsize
\setlength{\tabcolsep}{2.5pt}

\begin{minipage}{0.49\textwidth}
\centering
\textbf{(a) Precision}\\
\vspace{2pt}
\resizebox{\linewidth}{!}{
\begin{tabular}{lccccccccc}
\toprule
\textbf{Bench} & \textbf{Go} & \textbf{DL} & \textbf{CI} & \textbf{RF} & \textbf{LR} & \textbf{CNN} & \textbf{LSTM} & \textbf{Trans.} & \textbf{Ours} \\
\midrule
HP & 0.62 & 0.53 & 0.59 & 0.60 & 0.55 & 0.59 & 0.51 & 0.51 & \textbf{0.78} \\
HJ & 0.64 & \textbf{1.00} & 0.51 & 0.53 & 0.53 & 0.63 & 0.54 & 0.52 & 0.63 \\
MP & 0.55 & 0.54 & 0.67 & 0.66 & 0.64 & 0.58 & 0.53 & 0.56 & \textbf{0.71} \\
MJ & 0.63 & 0.28 & 0.58 & 0.58 & 0.56 & 0.60 & \textbf{0.76} & 0.54 & 0.66 \\
NP & 0.57 & 0.00 & 0.50 & \textbf{1.00} & 0.86 & 0.68 & 0.67 & 0.56 & 0.77 \\
NJ & 0.63 & 0.00 & 0.58 & 0.56 & 0.53 & 0.67 & \textbf{0.68} & 0.23 & 0.61 \\
EB & 0.58 & 0.00 & 0.44 & 0.36 & 0.40 & 0.64 & 0.62 & 0.54 & \textbf{0.67} \\
CE & 0.72 & 0.58 & 0.48 & 0.53 & 0.56 & 0.69 & \textbf{0.90} & 0.59 & 0.58 \\
\midrule
AVG & 0.62 & 0.37 & 0.54 & 0.60 & 0.58 & 0.64 & 0.65 & 0.51 & \textbf{0.67} \\
\bottomrule
\end{tabular}
}
\end{minipage}
\hfill
\begin{minipage}{0.49\textwidth}
\centering
\textbf{(b) Recall}\\
\vspace{2pt}
\resizebox{\linewidth}{!}{
\begin{tabular}{lccccccccc}
\toprule
\textbf{Bench} & \textbf{Go} & \textbf{DL} & \textbf{CI} & \textbf{RF} & \textbf{LR} & \textbf{CNN} & \textbf{LSTM} & \textbf{Trans.} & \textbf{Ours} \\
\midrule
HP & 0.66 & 0.07 & 0.61 & 0.54 & 0.57 & 0.59 & 0.50 & 0.51 & \textbf{0.77} \\
HJ & 0.57 & 0.02 & 0.79 & 0.71 & \textbf{0.86} & 0.60 & 0.51 & 0.52 & 0.79 \\
MP & 0.56 & 0.42 & 0.21 & 0.24 & 0.24 & \textbf{0.57} & 0.51 & 0.56 & 0.57 \\
MJ & 0.62 & 0.01 & 0.73 & 0.75 & \textbf{0.83} & 0.60 & 0.51 & 0.54 & 0.63 \\
NP & 0.51 & 0.00 & 0.08 & 0.50 & 0.50 & 0.58 & \textbf{0.67} & 0.53 & 0.57 \\
NJ & 0.51 & 0.00 & \textbf{0.92} & 0.83 & 0.83 & 0.62 & 0.69 & 0.50 & 0.71 \\
EB & 0.58 & 0.00 & 0.34 & 0.21 & 0.30 & 0.59 & \textbf{0.61} & 0.53 & 0.52 \\
CE & 0.64 & 0.24 & 0.82 & \textbf{0.94} & 0.82 & 0.67 & 0.89 & 0.55 & 0.66 \\
\midrule
AVG & 0.58 & 0.09 & 0.56 & 0.59 & 0.62 & 0.60 & 0.61 & 0.53 & \textbf{0.65} \\
\bottomrule
\end{tabular}
}
\end{minipage}

\vspace{0.8em}

\begin{minipage}{0.49\textwidth}
\centering
\textbf{(c) MCC}\\
\vspace{2pt}
\resizebox{\linewidth}{!}{
\begin{tabular}{lccccccccc}
\toprule
\textbf{Bench} & \textbf{Go} & \textbf{DL} & \textbf{CI} & \textbf{RF} & \textbf{LR} & \textbf{CNN} & \textbf{LSTM} & \textbf{Trans.} & \textbf{Ours} \\
\midrule
HP & 0.25 & 0.01 & 0.18 & 0.18 & 0.11 & 0.19 & 0.01 & 0.02 & \textbf{0.55} \\
HJ & 0.26 & 0.10 & 0.04 & 0.08 & 0.13 & 0.22 & 0.03 & 0.04 & \textbf{0.34} \\
MP & 0.10 & 0.07 & 0.14 & 0.15 & 0.14 & 0.15 & 0.04 & 0.12 & \textbf{0.34} \\
MJ & 0.26 & -0.05 & 0.21 & 0.20 & 0.20 & 0.21 & 0.09 & 0.08 & \textbf{0.30} \\
NP & 0.13 & 0.00 & 0.00 & \textbf{0.58} & 0.46 & 0.25 & 0.35 & 0.08 & 0.41 \\
NJ & 0.23 & -0.09 & 0.31 & 0.19 & 0.10 & 0.29 & \textbf{0.37} & 0.00 & 0.24 \\
EB & 0.16 & 0.00 & -0.08 & -0.18 & -0.14 & 0.23 & 0.24 & 0.07 & \textbf{0.28} \\
CE & 0.38 & 0.09 & -0.08 & 0.18 & 0.20 & 0.35 & \textbf{0.79} & 0.14 & 0.22 \\
\midrule
AVG & 0.22 & 0.02 & 0.09 & 0.17 & 0.15 & 0.24 & 0.24 & 0.07 & \textbf{0.33} \\
\bottomrule
\end{tabular}
}
\end{minipage}
\hfill
\begin{minipage}{0.49\textwidth}
\centering
\textbf{(d) AUC}\\
\vspace{2pt}
\resizebox{\linewidth}{!}{
\begin{tabular}{lccccccccc}
\toprule
\textbf{Bench} & \textbf{Go} & \textbf{DL} & \textbf{CI} & \textbf{RF} & \textbf{LR} & \textbf{CNN} & \textbf{LSTM} & \textbf{Trans.} & \textbf{Ours} \\
\midrule
HP & 0.66 & 0.53 & 0.56 & 0.61 & 0.57 & 0.66 & 0.52 & 0.55 & \textbf{0.82} \\
HJ & 0.68 & 0.59 & 0.57 & 0.58 & 0.61 & 0.66 & 0.53 & 0.55 & \textbf{0.75} \\
MP & 0.58 & 0.53 & 0.57 & 0.59 & 0.60 & 0.60 & 0.52 & 0.57 & \textbf{0.74} \\
MJ & 0.67 & 0.47 & 0.66 & 0.64 & 0.67 & 0.65 & 0.51 & 0.55 & \textbf{0.71} \\
NP & 0.61 & 0.54 & 0.72 & 0.75 & 0.75 & 0.62 & 0.72 & 0.59 & \textbf{0.77} \\
NJ & 0.66 & 0.57 & 0.72 & 0.60 & 0.70 & 0.74 & \textbf{0.80} & 0.63 & 0.72 \\
EB & 0.61 & 0.55 & 0.44 & 0.41 & 0.42 & 0.67 & 0.69 & 0.52 & \textbf{0.70} \\
CE & 0.78 & 0.56 & 0.53 & 0.59 & 0.56 & 0.73 & \textbf{0.98} & 0.62 & 0.67 \\
\midrule
AVG & 0.66 & 0.54 & 0.60 & 0.60 & 0.61 & 0.67 & 0.66 & 0.57 & \textbf{0.73} \\
\bottomrule
\end{tabular}
}
\end{minipage}

\vspace{2pt}
\begin{minipage}{0.98\textwidth}
\end{minipage}

\end{table*}

\begin{figure*}[!t]
\centering
\includegraphics[width=0.95\textwidth]{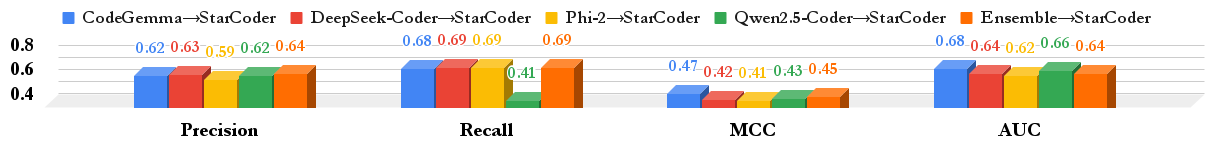}
\caption{The performance of CGMIA on detecting \textcolor{blue}{known} leaked APPS samples in StarCoder’s training data.}
\label{fig:apps}
\end{figure*}

\begin{table}[!t]
\caption{The performance of random forest and Gotcha with different LLMs as target and shadow models.}
    \centering
    \resizebox{\linewidth}{!}{
    \begin{tabular}{c|c|c|c|c|c|c|c|c|c|c|c|c}
    \hline
        \textbf{Metrics} &   \textbf{C→Q}  &   \textbf{D→Q}  &   \textbf{P→Q}  &   \textbf{Q→C}  &   \textbf{D→C}  &   \textbf{P→C}  &   \textbf{Q→D}  &   \textbf{C→D}  &   \textbf{P→D}  &   \textbf{C→P}  &   \textbf{D→P}  &   \textbf{Q→P}  \\ \hline
        \textbf{Precision}  & 0.77/0.65* & 0.72/0.65 & 0.58/0.55* & 0.78/0.66 & 0.69/0.66 & 0.62/0.57 & 0.74/0.66* & 0.77/0.65* & 0.59/0.56* & 0.78/0.56* & 0.76/0.56 & 0.76/0.55 \\ \hline
          \textbf{Recall}  & 0.74/0.65 & 0.57*/0.65 & 0.76/0.63 & 0.55/0.64 & 0.39*/0.62 & 0.69*/0.52 & 0.61/0.64 & 0.63/0.64 & 0.68/0.57 & 0.56/0.51 & 0.68*/0.50 & 0.54/0.49 \\ \hline
          \textbf{MCC}  & 0.56/0.30* & 0.44/0.30* & 0.76/0.63 & 0.47/0.31 & 0.39/0.30* & 0.69*/0.52 & 0.46/0.31 & 0.53/0.29* & 0.68/0.57 & 0.56/0.51 & 0.68*/0.50 & 0.54/0.49 \\ \hline
          \textbf{AUC}  & 0.81/0.70* & 0.76/0.70 & 0.60/0.57* & 0.79/0.70 & 0.72/0.70 & 0.65/0.59 & 0.78/0.71 & 0.82/0.70* & 0.60/0.58 & 0.74/0.58* & 0.77/0.57 & 0.73/0.57 \\ \hline
        
    \end{tabular}
    }
\label{tab:rfandgotcha}
\end{table}

% \begin{tcolorbox}[
%     colback=gray!5,      % Background color
%     colframe=gray!50,    % Border color
%     coltitle=black,      % Title text color
%     boxsep=0.01pt,       % Space between box and text
%     left=3pt,             % Left padding
%     right=3pt,            % Right padding
%     rounded corners,      % Rounded corners for the box
%     fonttitle=\bfseries        % Bold title font
% ]
% \textbf{Answer to RQ6:} CGMIA can identify more than 65\% of the 108 leaked APPS samples in StarCoder-7B’s training data, showing its effectiveness at detecting real-world benchmark leakage.
% \end{tcolorbox}

\subsection{RQ5: Can CGMIA work in StarCoder-based contamination-checkable and detected known-leakage settings?}\label{sec:rq5}

In our main experiments, we use four open-source code LLMs, including Qwen2.5-Coder, CodeGemma, DeepSeek-Coder, and Phi-2, and construct simulated benchmark leakage settings by applying LoRA fine-tuning on the selected eight benchmarks: HumanEval-X(Python), HumanEval-X(Java), MBXP-Python, MBXP-Java, NaturalCodeBench-Python, NaturalCodeBench-Java, EvoCode-Bench, and ClassEval. However, the technical reports of these models do not explicitly state whether all eight benchmarks used in this study were comprehensively decontaminated from their pre-training corpora. Therefore, it remains possible that some benchmark samples may have already appeared during pre-training, which could introduce contamination-induced bias and add noise to the reported measurements.

To further reduce this potential threat to validity, we additionally extend our experiments to StarCoder models, for which the Data Portraits~\cite{marone2023dataportraits} tool enables approximate verification of whether benchmark samples are present in the pre-training corpus. Data Portraits records training data using Bloom-filter-based sketches constructed from hashed strided $n$-grams and supports membership-style checking by testing whether query $n$-grams appear in the stored sketches~\cite{marone2023dataportraits}. Before constructing the controlled leakage setting, we first use the Data Portraits interface for the StarCoder-3B training corpus to examine all eight selected benchmarks and do not observe evidence of benchmark contamination. Based on this pre-check, we then construct controlled leakage settings on StarCoder-3B following the same experimental protocol as RQ1, where StarCoder-3B serves as both the target model and the shadow model for evaluating CGMIA.

Table~\ref{tab:starcoder_all_baselines} reports the Precision, Recall, MCC, and AUC results of CGMIA and all baseline methods across the eight code generation benchmarks. The benchmark and baseline abbreviations are consistent with those used in RQ1, and boldface indicates the best-performing method for each benchmark. As shown in Table~\ref{tab:starcoder_all_baselines}, CGMIA achieves the best average performance among all methods across all four evaluation metrics, obtaining a Precision of 0.67, Recall of 0.65, MCC of 0.33, and AUC of 0.73. Compared with the strongest baseline on each metric, CGMIA achieves absolute improvements of 0.02 in Precision, 0.03 in Recall, 0.09 in MCC, and 0.06 in AUC. Furthermore, CGMIA achieves the best MCC on five out of the eight benchmarks and the best AUC on six out of the eight benchmarks, indicating that CGMIA still maintains strong leakage detection capability on StarCoder-3B after Data Portraits-based contamination checking.

Similar to prior setups~\cite{gotcha,codemi}, our main experiments focus on simulated benchmark leakage, where LLMs are fine-tuned on subsets of benchmark data and the resulting fine-tuned models are treated as target models. To further evaluate CGMIA under a more realistic pre-training leakage scenario, we additionally conduct experiments on StarCoder-7B. Zhou et al.~\cite{LessLeak-Bench} confirmed that the pre-training corpus of StarCoder-7B contains 108 leaked samples from the APPS benchmark, making it a natural and practical target model for evaluating benchmark leakage detection in real-world settings. To construct the test set, we include two groups of samples. The first group consists of the 108 known leaked APPS samples identified by Zhou et al.~\cite{LessLeak-Bench} as appearing in StarCoder-7B's pre-training data. The second group consists of 108 APPS samples randomly selected from the remaining APPS dataset after excluding the known leaked samples, which are treated as samples with unconfirmed leakage status. We then query StarCoder-7B to generate outputs for all samples in the test set. 
To construct the shadow-model training data, we randomly select 108 samples from the remaining APPS dataset as member samples, representing data observed during shadow-model training, and another 108 samples as non-member samples, representing unseen data. This process is repeated five times to mitigate the influence of data randomness.

Figure~\ref{fig:apps} displays the average results of CGMIA's detection performance, employing four individual shadow models and an ensemble learning strategy. The results suggest that CGMIA can effectively recover known leaked benchmark samples in this realistic pre-training leakage scenario. Individual shadow models achieve Precision between 0.59 and 0.64, Recall between 0.65 and 0.69, MCC between 0.41 and 0.47, and AUC between 0.62 and 0.65. Notably, the Recall values indicate that CGMIA identifies 65\%--69\% of the 108 known leaked APPS samples in StarCoder's training data, demonstrating its practical utility. Moreover, the ensemble learning strategy achieves the highest Recall compared to all individual shadow models. However, since the remaining APPS samples cannot be strictly guaranteed to be free from undiscovered leakage, the reported Precision, MCC, and AUC results may still be affected by potential label noise in the samples with unconfirmed leakage status. Therefore, the results in this RQ should primarily be interpreted as evaluating CGMIA's ability to recover known leaked benchmark samples in a realistic pre-training leakage scenario.
\begin{tcolorbox}[
    colback=gray!5,
    colframe=gray!50,
    coltitle=black,
    boxsep=0.01pt,
    left=3pt,
    right=3pt,
    rounded corners,
    fonttitle=\bfseries
]
\textbf{Answer to RQ5:} CGMIA remains effective in both contamination-checkable and real-world known-leakage settings on StarCoder models, and can successfully recover a substantial proportion of detecting known leaked benchmark samples. 
\end{tcolorbox}

\begin{table}[!t]
\caption{The performance of CGMIA using the proposed ensemble learning strategy.}

    \centering
    \resizebox{\linewidth}{!}{
    \begin{tabular}{c|c|c|c|c|c|c|c|c|c|c|c|c}
    \hline
        \textbf{Metrics} &   \textbf{C/D/P→Q}  & \textbf{D→Q} & \textbf{C→Q} &   \textbf{C/P/Q→D}  & \textbf{P→D} & \textbf{C→D} &   \textbf{D/Q/P→C}  & \textbf{Q→C} & \textbf{P→C} &   \textbf{C/D/Q→P}  & \textbf{C→P} & \textbf{Q→P} \\ \hline
        \textbf{Precision}  & 0.74 & 0.73 & \textbf{0.81} & \textbf{0.84} & 0.71 & 0.79 & \textbf{0.77} & 0.73 & 0.74 & \textbf{0.78} & 0.71 & 0.69 \\ \hline
          \textbf{Recall}  & \textbf{0.82} & 0.73 & 0.77 & \textbf{0.83} & 0.64 & 0.77 & 0.72 & \textbf{0.74} & 0.62 & 0.71 & 0.67 & \textbf{0.78} \\ \hline
          \textbf{MCC}  & 0.49 & 0.45 & \textbf{0.57} & \textbf{0.67} & 0.44 & 0.57 & \textbf{0.50} & 0.45 & 0.49 & \textbf{0.48} & 0.41 & 0.43 \\ \hline
          \textbf{AUC}  & 0.81 & 0.81 & \textbf{0.87} & \textbf{0.9} & 0.86 & 0.87 & 0.82 & 0.81 & \textbf{0.90} & \textbf{0.80} & 0.72 & \textbf{0.80} \\ \hline
        
    \end{tabular}
    }
\label{tab:ensemble}
\end{table}

\subsection{\textbf{RQ6: Is CGMIA effective under imbalanced member sample ratios?}}
\label{sec:rq6}

\begin{figure}[t]
    \centering
    \includegraphics[width=\textwidth]{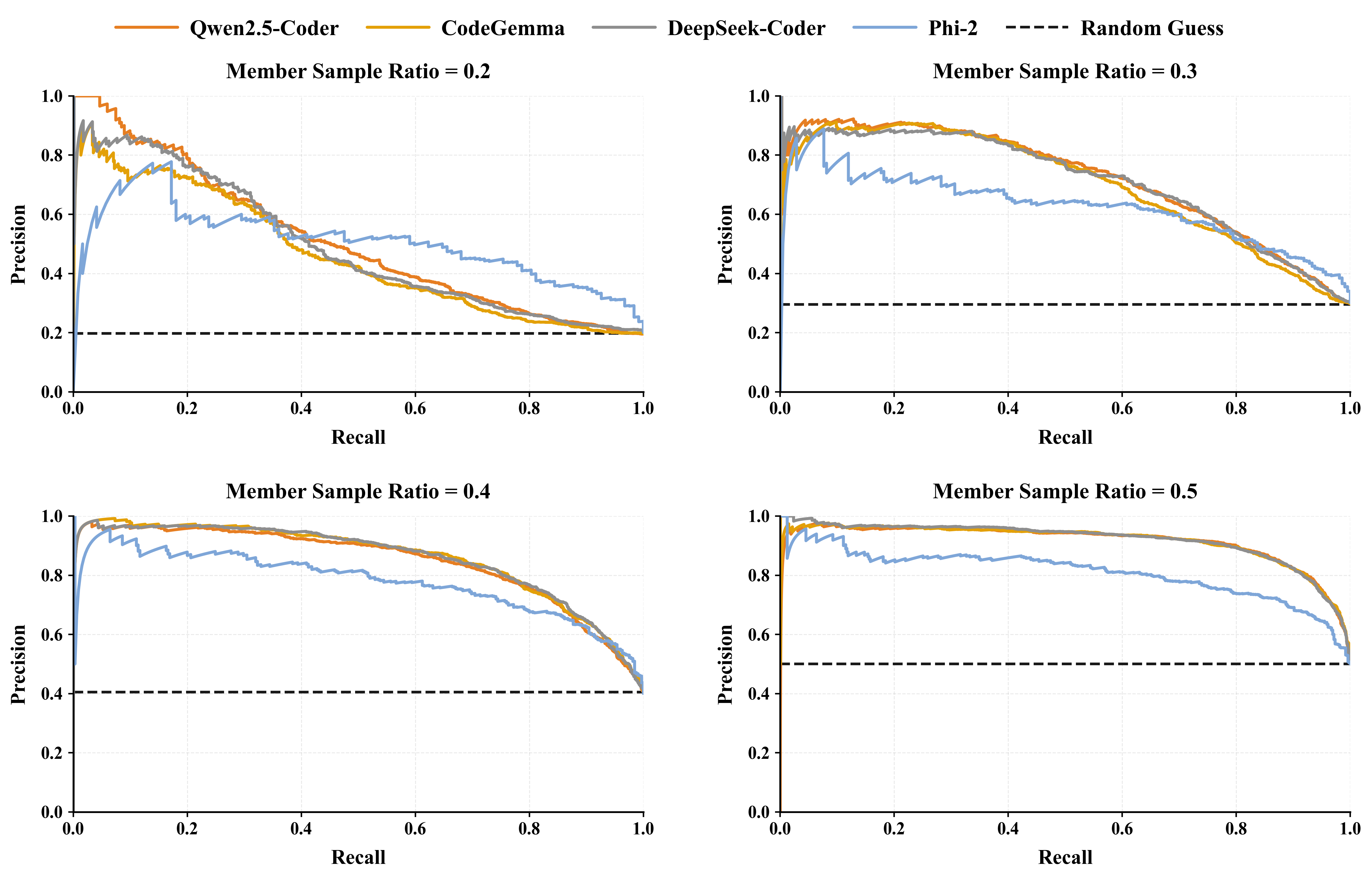}
    \caption{The precision-recall curves of CGMIA under different member ratios.}
    \label{fig:imbalanced_pr}
\end{figure}

In our previous experiments, we follow the common evaluation protocol adopted in prior MIA studies, including Shokri et al. \cite{shokri2017membership}, Buzzer \cite{buzzer}, and Hu et al. ~\cite{hu2022membership}, where member samples account for 50\% of the overall evaluation set. However, in practical benchmark leakage detection scenarios, leaked samples may constitute only a small fraction of all evaluated samples. Therefore, in this RQ, we further evaluate CGMIA under more imbalanced settings, where the proportions of member samples are reduced to 20\%, 30\%, and 40\%. 
We do not consider even lower member proportions because some benchmark datasets used in our study contain relatively limited numbers of samples. For example, NaturalCodeBench contains only 70 samples and HumanEval contains 164 samples. Under extremely low-prevalence settings, the number of member samples would become too small to support stable and statistically meaningful evaluation results. Figure~\ref{fig:imbalanced_pr} presents the Precision‑Recall (PR) curves of CGMIA under these settings, where recall is shown on the x‑axis and precision on the y‑axis. The black dashed horizontal line in each subplot denotes the naive random classifier baseline, whose precision value is equal to the proportion of member samples in the dataset.

As shown in Figure~\ref{fig:imbalanced_pr}, the detection performance of CGMIA generally declines as the member ratio decreases. This trend is expected because precision is sensitive to the prevalence of positive samples: when leaked samples become scarcer, achieving higher recall typically introduces more false positives, resulting in a more noticeable reduction in precision. Nevertheless, CGMIA still maintains meaningful discriminative capability across all evaluated ratios, as all PR curves remain substantially above their corresponding random-guess baselines. 
In particular, when the member ratio is 30\% or 40\%, Qwen2.5-Coder, CodeGemma, and DeepSeek-Coder maintain relatively high precision across a wide recall range. By contrast, Phi-2 exhibits comparatively weaker performance under these settings, likely due to its relatively limited code generation capability compared with the other target models. Even under the more challenging 20\% member-ratio setting, the PR curves of all evaluated models still remain substantially above the random-guess baseline, indicating that CGMIA continues to provide informative leakage detection signals under imbalanced settings.

\begin{tcolorbox}[
    colback=gray!5,
    colframe=gray!50,
    coltitle=black,
    boxsep=0.01pt,
    left=3pt,
    right=3pt,
    rounded corners,
    fonttitle=\bfseries
]
\textbf{Answer to RQ6:} CGMIA remains effective under imbalanced member-ratio settings and consistently achieves substantially better performance than random guessing.
\end{tcolorbox}

\subsection{Case Study}

\begin{figure}[!t]
    \centering
    \begin{minipage}[c]{0.55\textwidth} % 左图所在minipage，居中对齐
        \centering
        \includegraphics[height=9cm]{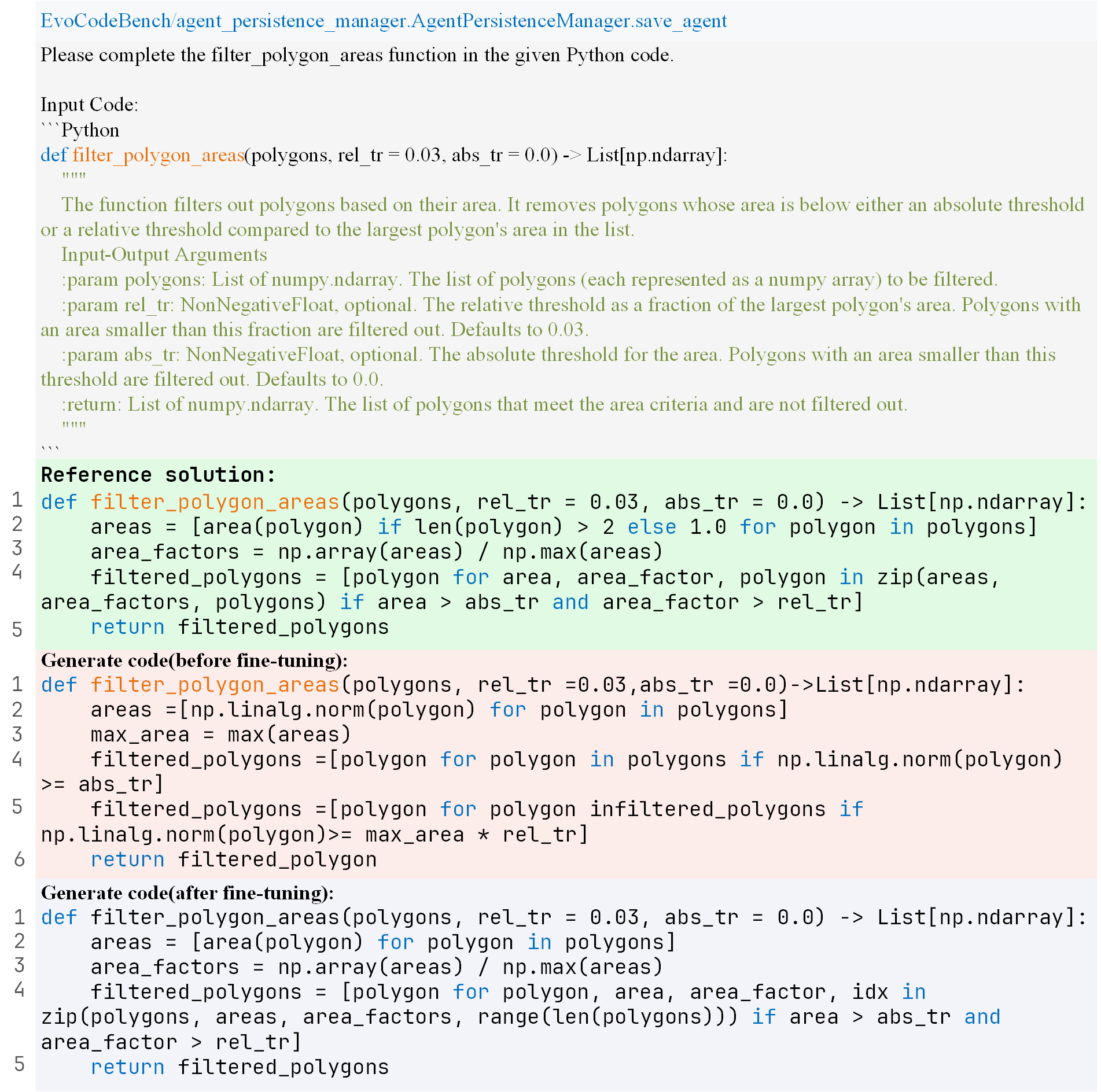} % 替换为左图实际路径
        \label{fig:sub1}
    \end{minipage}
    \hspace{0.08\textwidth} % 在两个minipage之间添加水平间距，可根据需要调整数值
    \begin{minipage}[c]{0.35\textwidth} % 右图所在minipage，居中对齐
        \vspace{-10pt}
        \centering
        \includegraphics[height=9cm]{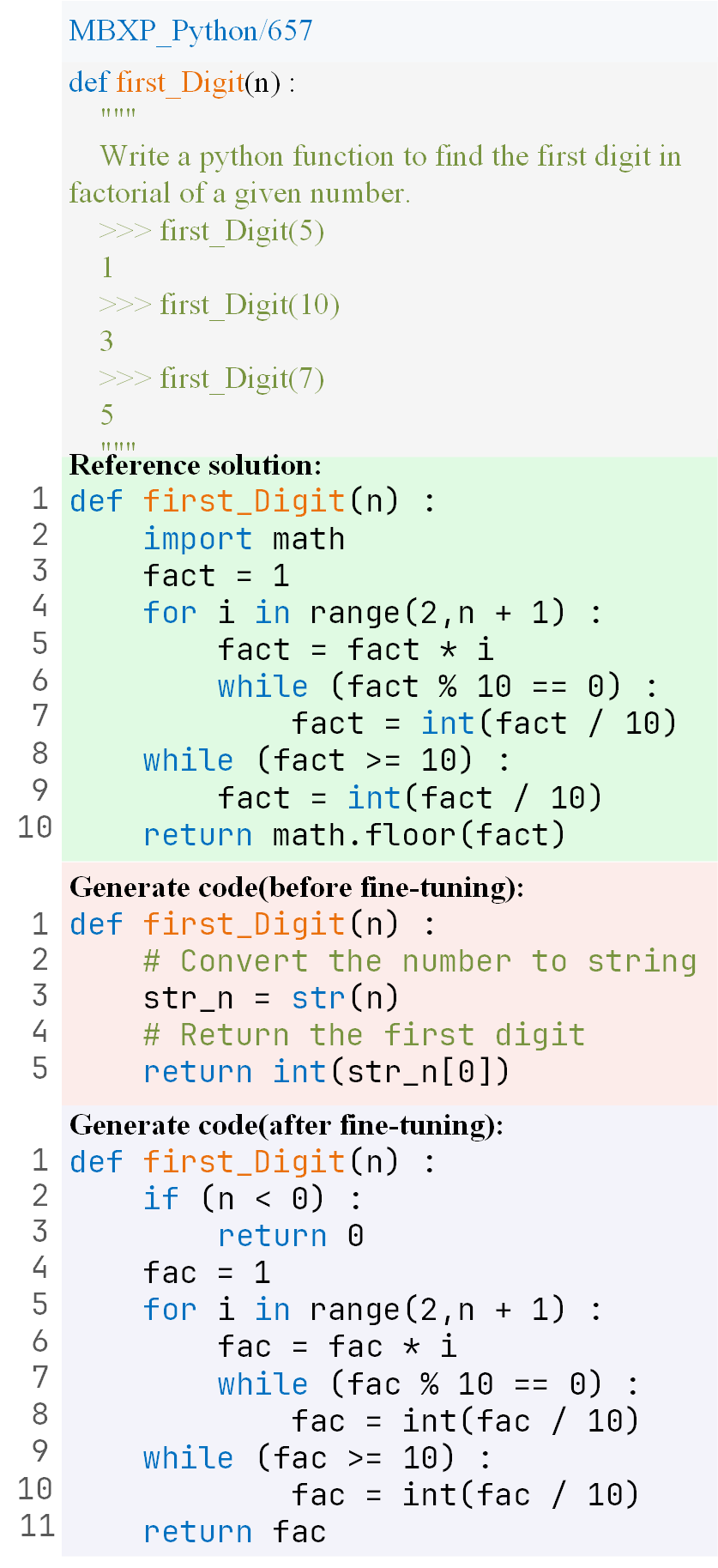} % 替换为右图实际路径
        \label{fig:sub2}
    \end{minipage}
    \vspace{-10pt}
    \caption{Samples that are successfully detected as leaked by our CGMIA method but not by baseline methods.}
    \label{fig:case_study}
    \vspace{-15pt}
\end{figure}

% 模型参数量，因为我们实验配置的问题只有一个4090的卡。 后面会继续探究大参数量的模型。 

% 只选了四个特征，有可能还有其他特征。 

% 我们是Python java的，有可能不适合其他语言。humaneval mbxp还有其他语言的，但是他们主要是专注于基础算法问题。没有其他比如c++这种专注真实开发场景的问题的被广泛使用的code generation benchmark。我们实验里面的几个都是Python java的benchmark。所以我们只探究了Python java的benchmark泄露。 We believe that this threat is minimal as our method is programming language agnostic and can be extended to other programming languages. We leave evaluation on other programming languages as future work and encourage other researchers to validate our findings on more datasets, including non-Python，Java datasets.

% 如果有人故意作弊，比如他 自己再去写一份不一样的编码的参考答案，然后基于这个参考答案训练llms。我们的方法有可能检测不出来，deserves a separate line of work, and we leave it for future work. 

Figure~\ref{fig:case_study} shows two examples from EvoCodeBench and MBXP-Python, with input prompts, reference solutions, and DeepSeek-Coder outputs before/after fine-tuning. Prior to fine-tuning, DeepSeek-Coder fails to produce correct solutions for both samples. After fine-tuning on one-quarter of the samples from these two benchmarks (including the two target samples, i.e., under data leakage conditions), the fine-tuned DeepSeek-Coder still generates incorrect code for the EvoCodeBench sample but successfully produces correct code for the MBXP-Python sample.

Compared to the pre-fine-tuning outputs, the code generated by the fine-tuned model more closely resembles the reference solutions. This observation provides an intuitive explanation of why membership inference can be effective for benchmark leakage detection. Once a benchmark sample has been exposed during training, the model does not always reproduce the reference solution exactly, but its generation behavior may become closer to the reference solution in terms of implementation logic, semantic structure, or functional behavior. Nevertheless, some discrepancies remain between the fine-tuned outputs and the references. Specifically, for the EvoCodeBench case, differences occur in lines 2 and 4; for the MBXP-Python case, the model replaces the variable name ``fact'' (used in the reference) with ``fac'' and introduces an additional conditional check in the generated code (lines 2--3). These two examples indicate that leakage signals in code generation are not limited to exact textual memorization. Instead, they may appear as partial similarity to the reference solution, improved functional correctness, or semantic closeness to the expected implementation.

Importantly, both samples are correctly identified as data leakage cases by our proposed CGMIA method. Among baseline approaches, only Gotcha successfully detects the EvoCodeBench sample, while all other baselines fail. For the MBXP-Python sample, none of the baseline methods detect the leakage. DetectLeak fails because it relies only on perplexity. When perplexity scores of the two generated code snippets are sorted in ascending order, these two samples do not fall into the lower-perplexity range used by DetectLeak to identify leaked samples, causing DetectLeak to misclassify these samples as non-leaked. This suggests that perplexity alone is insufficient, because a leaked sample may still have relatively high perplexity when the generated code contains uncommon implementation patterns or differs from the reference solution.

Although machine learning classifiers leverage four features (CodeBLEU, edit distance, perplexity, and test pass rate), the MBXP-Python sample has a test pass rate of 1 (a signal that should indicate data leakage). However, due to the relatively low similarity between its generated code and the reference solution, combined with a high perplexity score, the classifier still incorrectly labels the sample as non-leaked. This case shows that simply using expert features in a shallow classifier may be insufficient when different features provide inconsistent evidence. Baseline methods such as CNN, LSTM, and Transformer also miss these two leakage cases, and Gotcha specifically misses the MBXP-Python sample's leakage: they overemphasize semantic features while ignoring expert features---particularly the critical ``test pass rate'' signal of the MBXP-Python sample.

In contrast, CGMIA accurately identifies leakage in both samples because it integrates semantic embeddings with expert features. The semantic features capture the behavioral similarity between the generated code and the reference solution, while the expert features capture different observable effects of leakage, including reference similarity, likelihood familiarity, and functional correctness. Therefore, CGMIA does not depend on a single leakage signal being strong in all cases. For the EvoCodeBench sample, even though the generated code is still incorrect, the increased similarity to the reference solution provides useful evidence. For the MBXP-Python sample, even though the generated code is not textually identical to the reference solution, the successful test execution and semantic closeness provide complementary evidence. These cases help explain why CGMIA can work better than existing methods: benchmark leakage in code generation may leave multiple weak but complementary traces, and CGMIA is able to jointly capture these traces through the combination of expert features and semantic features.

\section{Threats of validity} 
\label{sec:Threats_of_validity}

(1) We conduct our experiments with two NVIDIA 4090 GPUs. Due to these hardware constraints, we limit our evaluation to smaller-scale models for fine-tuning as both target and shadow models. Consequently, the effectiveness of CGMIA on larger models with substantially more parameters remains uncertain and requires further investigation.

% (2) Our experimental code generation benchmarks encompass a wide spectrum of programming task difficulties, ranging from basic algorithmic problems to repository-level and class-level code generation in real-world development scenarios. Since the majority of benchmarks that reflect practical development environments are based on Python and Java, our experiments focus exclusively on these two languages. While we acknowledge that the results may not fully generalize to other programming languages such as C++ or JavaScript, we consider this limitation minimal, given that our method is programming language-agnostic and can be readily extended to other languages.
(2) We select only four expert-designed features in addition to semantic embeddings for membership inference. Other relevant features may exist that improve detection accuracy. Exploring additional or alternative features is an important direction for future research.  

(3) CGMIA relies on the key assumption that leaked training samples closely align with the reference solutions in code generation benchmarks. It specifically targets the scenario of unintentional data leakage in LLMs, typically involving the leakage of complete code segments. This aligns with findings from Zhou et al. \cite{LessLeak-Bench}, who note that LLMs primarily exhibit such unintentional data leakage. However, if an adversary deliberately modifies or rewrites reference code before training or fine-tuning the target LLM, CGMIA may fail to detect such deliberate leakage. Addressing this type of deliberate evasion tactic requires dedicated methodologies, which we reserve as future work.

(4) Our main experiments simulate leakage by applying LoRA fine-tuning on benchmark subsets, rather than faithfully reproducing incidental contamination in large-scale pre-training corpora. Compared with natural pre-training contamination, this controlled setup may induce stronger and more concentrated memorization signals because benchmark samples receive more direct exposure and are optimized through a different training process. As a result, membership inference may become easier, and the reported performance may be optimistic with respect to real-world pre-training leakage.
We adopt this design because it provides a controllable and reproducible MIA evaluation setting with known member and non-member labels, which are typically unavailable for closed-source or partially disclosed LLMs. Moreover, this practice is consistent with prior shadow-model-based MIA studies in the code domain, such as Gotcha \cite{gotcha}, CodeMI \cite{codemi}, and AdvPrompt-MIA \cite{AdvPrompt-MIA}, which commonly rely on controllable retraining or fine-tuning to construct labeled attack datasets for quantitative evaluation. Therefore, the primary goal of our experiments is to enable controlled and comparable evaluation of different leakage detection methods. In addition, the StarCoder/APPS experiment in Section \ref{sec:rq5} provides preliminary evidence under a known real leakage scenario. Since publicly verifiable cases of real pre-training contamination remain limited, evaluating CGMIA under broader real-world leakage settings remains important future work.

(5) Although our main experiments follow the balanced member/non-member setting commonly adopted in prior MIA studies, real-world benchmark leakage detection scenarios may involve much lower and unknown leakage prevalence. We address this issue in RQ6 by further conducting PR-curve analysis under different member ratios. The results show that CGMIA still maintains strong detection performance when the member proportion is reduced to 20\%, 30\%, and 40\%. Moreover, unlike threshold-dependent methods such as DetectLeak, which suffer from threshold sensitivity due to prevalence-dependent threshold selection, CGMIA does not require manually specifying a membership threshold, making it more practical under unknown-prevalence settings.

(6) CGMIA assumes a score-access black-box interface rather than a strict text-only interface. Specifically, the perplexity feature requires token-level output scores such as log-probabilities. Although such score information is available in many practical LLM systems, including commercial models such as OpenAI GPT and Gemini, as well as open-weight models deployed through modern inference frameworks, it is not guaranteed to be accessible in all commercial black-box LLM services. Therefore, the current version of CGMIA is applicable to score-access black-box settings, but not directly to strict text-only APIs that return only generated code.

(7) In CGMIA, CodeBERT-based semantic representation may still introduce risks for long code sequences. 
To avoid direct truncation, we adopt a sliding-window encoding strategy with average pooling for code longer than
CodeBERT's maximum input length. However, this strategy may still lose fine-grained long-range
dependencies or dilute localized memorization signals when multiple chunks are averaged into
a single representation. As a result, the semantic features of very long generated programs or
reference solutions may be less precise. Future work could explore long-context code encoders or
structure-aware representations to better capture long-code semantics.

(8) In RQ5, the evaluation on StarCoder-7B may still be affected by label uncertainty in the negative samples. Although Zhou et al.~\cite{LessLeak-Bench} confirmed 108 detecting leaked APPS samples in the StarCoder-7B pre-training corpus, the remaining APPS samples cannot be strictly guaranteed to be free from undiscovered leakage. Therefore, the samples used as negatives in our evaluation are more accurately treated as samples with unconfirmed leakage status rather than verified non-leaked samples. As a result, the reported Precision, MCC, and AUC values may still be influenced by potential label noise in the evaluation set. To mitigate over-interpretation, we primarily frame RQ5 as evaluating CGMIA’s ability to recover detecting known leaked benchmark samples under a realistic pre-training leakage scenario, with particular emphasis on Recall for the confirmed leaked subset. Future work could further improve negative-sample verification through stronger contamination-detection heuristics or more direct access to pre-training corpus snapshots when available.

%本文实验通过在基准数据集子集上开展 LoRA 微调来模拟数据泄露，并未完整复现大规模预训练语料中偶然混入污染数据的真实场景。相较于自然场景下的预训练数据泄露，该受控实验设定会因样本曝光度更高、优化过程存在差异，产生更强且更集中的记忆特征，进而降低成员推理的检测难度，使得实验性能评估结果偏于乐观。我们采用该实验设计，并非旨在完全等价于真实场景中的预训练数据污染问题，而是为了构建成员 / 非成员标签完全已知的可控评估环境；对于训练数据未公开的闭源大语言模型而言，这类真实标签数据往往难以获取。因此，本文实验结论仅可证明该方法在可控泄露模拟场景下的有效性，针对存在真实数据污染的模型开展进一步验证，仍有待未来研究完善。%

\section{Conclusion}
\label{sec:Conclusion}

We propose CGMIA, an MIA method designed to detect potential data leakage in code generation benchmarks. CGMIA employs a shadow modeling strategy to generate labeled training data distinguishing members from non-members. It then extracts both expert features and semantic features from the data to build a classifier. Extensive experiments confirm that CGMIA not only outperforms existing MIA methods but also demonstrates strong practical effectiveness.  
This work not only advances methodologies for detecting leakage in code generation benchmarks but also highlights the value of integrating heterogeneous features, paving the way for future research into detecting data leakage in other types of benchmarks.

\section{Data Availability}

The benchmark data configurations, experimental code, and replication materials used in this study are publicly available at \url{https://github.com/Ricardo-J-Chen/CGMIA}. The repository serves as the online appendix for this paper and contains the benchmark data files, prompt templates, LoRA fine-tuning and code-generation scripts (e.g., \texttt{finetune/fintune.py} and \texttt{finetune/generate/generate\_all.py}), feature-extraction and membership inference scripts, CGMIA and baseline attack implementations (e.g., \texttt{MIA/cgmia.py}, \texttt{MIA/gotcha.py}, and \texttt{MIA/detectLeak.py}), ablation-study and ensemble-learning scripts (e.g., \texttt{MIA/ablation.py} and \texttt{MIA/Ensemble\_learning.py}), and the data-generation scripts for the tables and figures reported in the manuscript. All tables and figures reported in the manuscript can be reproduced using the code and materials provided in the repository.

\section{Acknowledgments} 
This research was supported by the National Natural Science Foundation of China under Grant No. 62502440 and the Zhejiang Provincial Natural Science Foundation of China under Grant No. LQN26F020003.

% 62502440 LQN26F020003  是余啸的基金号

\bibliographystyle{ACM-Reference-Format}

\bibliography{article/ref}

\appendix
\clearpage
\appendix

\section{Supplementary Machine Learning Baseline Results}
\label{app:supplementary_ml_baselines}

This appendix provides supplementary Precision, Recall, MCC, and AUC results for traditional machine learning baselines (SVM, NB, DT, MLP), complementing the LR and RF results shown in the main text.

\begin{table}[H]
\centering
\caption{The Precision results of other traditional machine learning baselines.}
\label{tab:app_ml_precision}
\scriptsize
\setlength{\tabcolsep}{3.2pt}
\renewcommand{\arraystretch}{0.82}

\begin{minipage}[t]{0.48\textwidth}
\centering
\textbf{(a) Qwen2.5-Coder}

\vspace{1mm}

\begin{tabular}{lcccc}
\toprule
Bench & SVM & NB & DT & MLP \\
\midrule
HP & 0.52 & 0.84 & \textbf{0.70} & \textbf{0.70} \\
HJ & 0.50 & 0.68 & \textbf{0.67} & 0.63 \\
MP & 0.50 & 0.59 & 0.51 & \textbf{0.54} \\
MJ & 0.50 & 0.52 & 0.36 & \textbf{0.37} \\
NP & 0.50 & 0.50 & -- & -- \\
NJ & 0.50 & \textbf{1.00} & \textbf{0.70} & \textbf{0.70} \\
EB & 0.51 & 0.68 & \textbf{0.52} & 0.49 \\
CE & 0.50 & 0.52 & \textbf{0.41} & 0.35 \\
AVG & 0.50 & 0.67 & \textbf{0.55} & 0.54 \\
\bottomrule
\end{tabular}
\end{minipage}
\hfill
\begin{minipage}[t]{0.48\textwidth}
\centering
\textbf{(b) CodeGemma}

\vspace{1mm}

\begin{tabular}{lcccc}
\toprule
Bench & SVM & NB & DT & MLP \\
\midrule
HP & 0.50 & 0.43 & \textbf{0.70} & 0.00 \\
HJ & 0.50 & 0.68 & \textbf{0.70} & 0.80 \\
MP & 0.50 & 0.69 & \textbf{0.55} & 0.76 \\
MJ & 0.50 & 0.59 & 0.42 & \textbf{0.83} \\
NP & 0.55 & -- & -- & -- \\
NJ & 0.50 & -- & \textbf{0.70} & -- \\
EB & 0.50 & 0.00 & \textbf{0.40} & -- \\
CE & 0.54 & \textbf{0.60} & 0.27 & -- \\
AVG & 0.51 & 0.50 & \textbf{0.53} & 0.60 \\
\bottomrule
\end{tabular}
\end{minipage}

\vspace{2mm}

\begin{minipage}[t]{0.48\textwidth}
\centering
\textbf{(c) DeepSeek-Coder}

\vspace{1mm}

\begin{tabular}{lcccc}
\toprule
Bench & SVM & NB & DT & MLP \\
\midrule
HP & 0.68 & 0.93 & \textbf{0.70} & 0.79 \\
HJ & 0.51 & 0.65 & \textbf{0.70} & \textbf{1.00} \\
MP & 0.50 & 0.56 & \textbf{0.56} & 0.59 \\
MJ & 0.50 & 0.50 & \textbf{0.45} & 0.66 \\
NP & \textbf{0.50} & -- & -- & 0.25 \\
NJ & 0.50 & \textbf{0.91} & -- & 0.55 \\
EB & 0.49 & 0.51 & \textbf{0.70} & 0.49 \\
CE & 0.52 & 0.50 & \textbf{0.34} & 0.50 \\
AVG & 0.53 & 0.65 & \textbf{0.57} & 0.60 \\
\bottomrule
\end{tabular}
\end{minipage}
\hfill
\begin{minipage}[t]{0.48\textwidth}
\centering
\textbf{(d) Phi-2}

\vspace{1mm}

\begin{tabular}{lcccc}
\toprule
Bench & SVM & NB & DT & MLP \\
\midrule
HP & 0.68 & \textbf{0.80} & \textbf{0.80} & \textbf{0.80} \\
HJ & 0.56 & 0.70 & 0.73 & \textbf{0.80} \\
MP & 0.41 & 0.53 & 0.56 & \textbf{0.65} \\
MJ & 0.33 & 0.39 & 0.40 & \textbf{0.51} \\
NP & 0.32 & 0.40 & 0.51 & \textbf{0.62} \\
NJ & 0.41 & 0.41 & 0.59 & \textbf{0.62} \\
EB & 0.58 & 0.64 & 0.66 & \textbf{0.68} \\
CE & 0.44 & 0.42 & \textbf{0.51} & 0.48 \\
AVG & 0.47 & 0.54 & 0.60 & \textbf{0.65} \\
\bottomrule
\end{tabular}
\end{minipage}

\end{table}

\begin{table}[H]
\centering
\caption{The Recall values of additional traditional machine learning baselines.}
\label{tab:app_ml_recall}
\scriptsize
\setlength{\tabcolsep}{3.2pt}
\renewcommand{\arraystretch}{0.82}

\begin{minipage}[t]{0.48\textwidth}
\centering
\textbf{(a) Qwen2.5-Coder}

\vspace{1mm}

\begin{tabular}{lcccc}
\toprule
Bench & SVM & NB & DT & MLP \\
\midrule
HP & \textbf{0.70} & 0.63 & 0.50 & 0.57 \\
HJ & 0.70 & 0.70 & 1.00 & 1.00 \\
MP & \textbf{0.70} & 0.68 & 0.74 & 0.67 \\
MJ & \textbf{0.70} & \textbf{0.70} & 0.98 & 0.98 \\
NP & \textbf{0.70} & \textbf{0.70} & 0.00 & 0.00 \\
NJ & \textbf{0.70} & 0.53 & 0.42 & 0.75 \\
EB & \textbf{0.70} & \textbf{0.70} & 0.79 & 0.94 \\
CE & 0.35 & 0.64 & 0.88 & \textbf{1.00} \\
AVG & \textbf{0.66} & \textbf{0.66} & 0.66 & 0.74 \\
\bottomrule
\end{tabular}
\end{minipage}
\hfill
\begin{minipage}[t]{0.48\textwidth}
\centering
\textbf{(b) CodeGemma}

\vspace{1mm}

\begin{tabular}{lcccc}
\toprule
Bench & SVM & NB & DT & MLP \\
\midrule
HP & \textbf{0.70} & 0.11 & 0.11 & 0.00 \\
HJ & \textbf{0.70} & 0.61 & \textbf{1.00} & 0.43 \\
MP & \textbf{0.70} & 0.25 & 0.54 & 0.17 \\
MJ & \textbf{0.70} & 0.46 & 0.98 & 0.14 \\
NP & \textbf{0.70} & 0.00 & 0.00 & 0.00 \\
NJ & \textbf{0.70} & 0.00 & 0.25 & 0.00 \\
EB & 0.64 & 0.00 & \textbf{1.00} & 0.00 \\
CE & 0.11 & 0.35 & \textbf{1.00} & 0.00 \\
AVG & \textbf{0.62} & 0.22 & 0.61 & 0.09 \\
\bottomrule
\end{tabular}
\end{minipage}

\vspace{2mm}

\begin{minipage}[t]{0.48\textwidth}
\centering
\textbf{(c) DeepSeek-Coder}

\vspace{1mm}

\begin{tabular}{lcccc}
\toprule
Bench & SVM & NB & DT & MLP \\
\midrule
HP & \textbf{0.70} & 0.59 & 0.43 & 0.38 \\
HJ & 0.70 & 0.70 & 1.00 & 0.70 \\
MP & \textbf{0.70} & 0.69 & 0.87 & 0.62 \\
MJ & \textbf{0.70} & \textbf{0.70} & 0.97 & \textbf{0.70} \\
NP & \textbf{0.53} & 0.30 & 0.00 & 0.13 \\
NJ & \textbf{0.70} & 0.53 & 0.00 & \textbf{0.70} \\
EB & \textbf{0.70} & \textbf{0.70} & 0.38 & \textbf{0.70} \\
CE & 0.58 & \textbf{0.70} & 0.82 & \textbf{0.70} \\
AVG & \textbf{0.66} & 0.54 & 0.56 & 0.55 \\
\bottomrule
\end{tabular}
\end{minipage}
\hfill
\begin{minipage}[t]{0.48\textwidth}
\centering
\textbf{(d) Phi-2}

\vspace{1mm}

\begin{tabular}{lcccc}
\toprule
Bench & SVM & NB & DT & MLP \\
\midrule
HP & \textbf{0.80} & 0.76 & 0.69 & 0.48 \\
HJ & \textbf{0.80} & 0.73 & 0.73 & 0.62 \\
MP & \textbf{0.77} & 0.71 & 0.63 & 0.48 \\
MJ & \textbf{0.80} & 0.78 & 0.76 & 0.72 \\
NP & 0.72 & \textbf{0.80} & \textbf{0.80} & 0.55 \\
NJ & 0.72 & 0.72 & 0.72 & 0.55 \\
EB & \textbf{0.69} & 0.57 & 0.48 & 0.12 \\
CE & 0.62 & \textbf{0.74} & 0.68 & 0.56 \\
AVG & \textbf{0.74} & 0.73 & 0.69 & 0.51 \\
\bottomrule
\end{tabular}
\end{minipage}

\end{table}

\begin{table}[H]
\centering
\caption{The MCC values of additional traditional machine learning baselines.}
\label{tab:app_ml_mcc}
\scriptsize
\setlength{\tabcolsep}{3.2pt}
\renewcommand{\arraystretch}{0.82}

\begin{minipage}[t]{0.48\textwidth}
\centering
\textbf{(a) Qwen2.5-Coder}

\vspace{1mm}

\begin{tabular}{lcccc}
\toprule
Bench & SVM & NB & DT & MLP \\
\midrule
HP & 0.19 & \textbf{0.75} & 0.58 & 0.63 \\
HJ & 0.00 & 0.60 & \textbf{0.96} & 0.93 \\
MP & 0.00 & 0.40 & \textbf{0.56} & \textbf{0.56} \\
MJ & 0.00 & 0.17 & 0.54 & \textbf{0.55} \\
NP & 0.00 & 0.00 & 0.00 & 0.00 \\
NJ & 0.00 & \textbf{0.85} & 0.51 & 0.77 \\
EB & 0.15 & 0.61 & 0.62 & \textbf{0.70} \\
CE & 0.00 & 0.10 & 0.54 & \textbf{0.55} \\
AVG & 0.04 & 0.44 & 0.54 & \textbf{0.59} \\
\bottomrule
\end{tabular}
\end{minipage}
\hfill
\begin{minipage}[t]{0.48\textwidth}
\centering
\textbf{(b) CodeGemma}

\vspace{1mm}

\begin{tabular}{lcccc}
\toprule
Bench & SVM & NB & DT & MLP \\
\midrule
HP & 0.00 & -0.05 & \textbf{0.24} & -0.13 \\
HJ & 0.00 & 0.32 & \textbf{1.00} & 0.36 \\
MP & 0.00 & 0.18 & \textbf{0.47} & 0.19 \\
MJ & 0.00 & 0.15 & \textbf{0.64} & 0.20 \\
NP & \textbf{0.30} & 0.00 & 0.00 & 0.00 \\
NJ & 0.00 & 0.00 & \textbf{0.38} & 0.00 \\
EB & 0.04 & -0.15 & \textbf{0.64} & 0.00 \\
CE & 0.06 & 0.13 & \textbf{0.37} & 0.00 \\
AVG & 0.05 & 0.07 & \textbf{0.47} & 0.08 \\
\bottomrule
\end{tabular}
\end{minipage}

\vspace{2mm}

\begin{minipage}[t]{0.48\textwidth}
\centering
\textbf{(c) DeepSeek-Coder}

\vspace{1mm}

\begin{tabular}{lcccc}
\toprule
Bench & SVM & NB & DT & MLP \\
\midrule
HP & 0.60 & \textbf{0.82} & 0.52 & 0.51 \\
HJ & 0.13 & 0.55 & \textbf{1.00} & \textbf{1.00} \\
MP & 0.00 & 0.34 & \textbf{0.73} & 0.34 \\
MJ & 0.00 & 0.09 & \textbf{0.68} & 0.56 \\
NP & \textbf{0.00} & \textbf{0.00} & \textbf{0.00} & -0.35 \\
NJ & 0.00 & \textbf{0.75} & 0.00 & 0.30 \\
EB & 0.00 & 0.18 & \textbf{0.49} & 0.00 \\
CE & 0.08 & 0.00 & \textbf{0.37} & 0.00 \\
AVG & 0.10 & 0.34 & \textbf{0.47} & 0.29 \\
\bottomrule
\end{tabular}
\end{minipage}
\hfill
\begin{minipage}[t]{0.48\textwidth}
\centering
\textbf{(d) Phi-2}

\vspace{1mm}

\begin{tabular}{lcccc}
\toprule
Bench & SVM & NB & DT & MLP \\
\midrule
HP & 0.67 & \textbf{0.76} & 0.70 & 0.52 \\
HJ & 0.52 & 0.62 & \textbf{0.66} & 0.63 \\
MP & 0.23 & \textbf{0.39} & 0.37 & 0.36 \\
MJ & 0.06 & 0.21 & 0.21 & \textbf{0.37} \\
NP & -0.07 & 0.25 & \textbf{0.44} & 0.39 \\
NJ & 0.18 & 0.18 & \textbf{0.48} & 0.39 \\
EB & \textbf{0.45} & 0.42 & 0.39 & 0.16 \\
CE & 0.17 & 0.22 & \textbf{0.34} & 0.21 \\
AVG & 0.28 & 0.38 & \textbf{0.45} & 0.38 \\
\bottomrule
\end{tabular}
\end{minipage}

\end{table}

\begin{table}[H]
\centering
\caption{The AUC values of additional traditional machine learning baselines.}
\label{tab:app_ml_auc}
\scriptsize
\setlength{\tabcolsep}{3.2pt}
\renewcommand{\arraystretch}{0.82}

\begin{minipage}[t]{0.48\textwidth}
\centering
\textbf{(a) Qwen2.5-Coder}

\vspace{1mm}

\begin{tabular}{lcccc}
\toprule
Bench & SVM & NB & DT & MLP \\
\midrule
HP & 0.79 & 0.65 & 0.60 & \textbf{0.66} \\
HJ & 0.80 & 0.64 & \textbf{0.70} & \textbf{0.70} \\
MP & 0.67 & \textbf{0.56} & 0.53 & \textbf{0.56} \\
MJ & 0.66 & 0.42 & 0.53 & \textbf{0.57} \\
NP & 0.56 & 0.26 & 0.20 & \textbf{0.34} \\
NJ & 0.61 & \textbf{0.67} & 0.62 & \textbf{0.67} \\
EB & 0.62 & \textbf{0.64} & 0.60 & 0.61 \\
CE & 0.54 & 0.40 & \textbf{0.51} & 0.46 \\
AVG & 0.66 & 0.53 & 0.54 & \textbf{0.57} \\
\bottomrule
\end{tabular}
\end{minipage}
\hfill
\begin{minipage}[t]{0.48\textwidth}
\centering
\textbf{(b) CodeGemma}

\vspace{1mm}

\begin{tabular}{lcccc}
\toprule
Bench & SVM & NB & DT & MLP \\
\midrule
HP & 0.50 & 0.49 & \textbf{0.38} & 0.51 \\
HJ & 0.50 & 0.68 & \textbf{0.70} & 0.66 \\
MP & 0.50 & 0.67 & \textbf{0.53} & 0.58 \\
MJ & 0.50 & 0.59 & \textbf{0.57} & 0.59 \\
NP & 0.50 & \textbf{0.59} & 0.20 & 0.50 \\
NJ & 0.50 & 0.51 & \textbf{0.66} & 0.50 \\
EB & 0.50 & 0.59 & \textbf{0.65} & 0.50 \\
CE & 0.50 & 0.52 & \textbf{0.46} & 0.50 \\
AVG & 0.50 & 0.58 & \textbf{0.52} & 0.54 \\
\bottomrule
\end{tabular}
\end{minipage}

\vspace{2mm}

\begin{minipage}[t]{0.48\textwidth}
\centering
\textbf{(c) DeepSeek-Coder}

\vspace{1mm}

\begin{tabular}{lcccc}
\toprule
Bench & SVM & NB & DT & MLP \\
\midrule
HP & 0.95 & \textbf{0.97} & 0.69 & 0.85 \\
HJ & 0.99 & 0.91 & \textbf{1.00} & \textbf{1.00} \\
MP & 0.81 & 0.80 & \textbf{0.89} & 0.63 \\
MJ & 0.86 & 0.73 & \textbf{0.90} & 0.82 \\
NP & \textbf{0.62} & 0.51 & 0.50 & 0.38 \\
NJ & 0.93 & \textbf{0.98} & 0.50 & 0.58 \\
EB & 0.61 & 0.64 & \textbf{0.70} & 0.50 \\
CE & 0.61 & 0.44 & \textbf{0.77} & 0.50 \\
AVG & \textbf{0.80} & 0.75 & 0.74 & 0.66 \\
\bottomrule
\end{tabular}
\end{minipage}
\hfill
\begin{minipage}[t]{0.48\textwidth}
\centering
\textbf{(d) Phi-2}

\vspace{1mm}

\begin{tabular}{lcccc}
\toprule
Bench & SVM & NB & DT & MLP \\
\midrule
HP & 0.80 & 0.80 & 0.80 & 0.80 \\
HJ & \textbf{0.79} & 0.78 & 0.77 & \textbf{0.79} \\
MP & \textbf{0.68} & 0.67 & 0.64 & 0.67 \\
MJ & \textbf{0.67} & 0.55 & 0.58 & \textbf{0.67} \\
NP & 0.63 & 0.64 & 0.63 & \textbf{0.67} \\
NJ & 0.59 & 0.66 & \textbf{0.70} & 0.67 \\
EB & \textbf{0.71} & 0.67 & 0.65 & 0.68 \\
CE & 0.55 & 0.51 & 0.60 & \textbf{0.63} \\
AVG & 0.68 & 0.66 & 0.67 & \textbf{0.70} \\
\bottomrule
\end{tabular}
\end{minipage}

\end{table}

\end{document}